\documentclass[11pt,titlepage]{article}

\usepackage{setspace}
\usepackage{amsmath}

\usepackage{graphicx}

\usepackage[round,numbers,sort&compress]{natbib}
\newcommand{\be}{\begin{equation}}
\newcommand{\ee}{\end{equation}}
\newcommand{\bea}{\begin{eqnarray}}
\newcommand{\eea}{\end{eqnarray}}

\title{Non-equilibrium self-assembly of a filament coupled to ATP/GTP hydrolysis}

\author{Padinhateeri Ranjith\thanks{pranjith@curie.fr,
       Tel.:~ (33) 156 246 472, Fax:~(33) 140 510 636}\\
        Physico-Chimie UMR 168\\
    Institut Curie, 26 rue d'Ulm \\
    75248 Paris Cedex 05, France
    \and David Lacoste \\
    Laboratoire de Physico-Chimie Th\'{e}orique, \\
    ESPCI, 10 rue Vauquelin, \\
    Paris Cedex 05, France\\
    \and Kirone Mallick\\
    Service de Physique Th\'{e}orique\\
     CEA Saclay\\
     91191 Gif, France
 \and Jean-Fran\c{c}ois Joanny \\
    Physico-Chimie UMR 168 \\
    Institut Curie, 26 rue d'Ulm \\
    75248 Paris Cedex 05, France
    }

\date{}

\begin{document}

% generate the title page from the info in the headers above
\maketitle

% 200 words max Abstract
\abstract{

We study the stochastic dynamics of growth and shrinkage of single
actin filaments or microtubules taking into account  insertion,
removal, and ATP/GTP hydrolysis of subunits. The resulting phase diagram
 contains three different phases: a rapidly growing
phase, an intermediate phase and a bound phase. We analyze all these phases, 
with an emphasis on the bound phase. We also
discuss how hydrolysis affects force-velocity
curves. The bound phase shows features of dynamic instability, which
we characterize in terms of the time needed for the ATP/GTP cap to
disappear as well as the time needed for the filament to reach a
length of zero ({\it i.e.} to collapse) for the first time. We
obtain exact expressions for all these quantities, which we test
using Monte Carlo simulations.

\emph{Key words:} Actin; Microtubule; Dynamic instability; ATP hydrolysis;
Force-velocity curves; Stochastic dynamics}

% New page
\clearpage

\section*{Introduction}
A large number of structural elements of cells are made of fibers.
Well studied examples of these fibers are microtubules and actin
filaments. Microtubules are able to undergo rapid dynamic
transitions between growth (polymerization) and decay
(depolymerization) in a process called dynamic instability 
\cite{Desai_Mitchison_MT:97}. Actin filaments are able to
undergo treadmilling-like motion. These dynamic features of
microtubules and actin filaments play an essential role in
cellular biology \cite{howard-book}. For instance, the treadmilling of actin
filaments occurs in filopodia, lamellipodia, flagella and
stereocilia \cite{Schneider:02,rosenbaum:02,Prost_Joanny_BPJ:07}. Actin growth dynamics
is also important in acrosome reactions, where sperm fuses with
egg~\cite{liu-etal:99,shin-etal:04,breitbart-etal:05}. During cell
division, the movements of chromosomes are coupled to the
elongation and shortening of the microtubules to which they
bind~\cite{hunt-etal:98, howard-book}.

Energy dissipation is critical for these dynamic non-equilibrium 
features of microtubules and actin.
Energy is dissipated when ATP (respectively GTP)
associated to actin monomers (respectively tubulin dimers) is
irreversibly hydrolyzed into ADP (respectively GDP). Since this hydrolysis
process typically lags behind the assembly process, a cap of
consecutive ATP/GTP subunits can form at the end of the filament
\cite{hill:85,hill:86}.

Let us first consider studies of the dynamic instability of microtubules.
Non-equilibrium properties of microtubules have often been
described using a phenomenological two-state model developed by Dogterom and Leibler
\cite{Leibler:93}. In such a model, a microtubule exists either in
a rescue phase (where a GTP cap exists at the end of the
microtubule) or a catastrophe phase (with no GTP cap), with
stochastic transitions between the two states. A limitation
of such a model is that a switching frequency is built in the
model rather than derived from a precise theoretical modeling of
the GTP cap. This question was addressed later by Flyvbjerg et al. 
\cite{Leibler-cap:94,Leibler-cap:96}, where a theory for the
dynamics of the GTP cap was included. At about the same time, a mathematical 
analysis of the Dogterom-Leibler model using Green functions formalism 
was carried out in ref.~\cite{Bicout_PRE:97}.
The study of Flyvbjerg et al. \cite{Leibler-cap:94,Leibler-cap:96}, was
generalized in \cite{Wolynes:06}, with the use of a variational method 
and numerical simulations. This kind of stochastic model for the dynamic instability 
of microtubules was further studied by Antal et al. 
 \cite{Antal-etal-PRE:07,Antal-etal-JSP:07}.
The Antal et al. model takes into account the addition and hydrolysis of GTP
subunits, and the removal of GDP subunits. Exact calculations are
carried out in some particular cases such as when the GDP
detachment rate goes to zero or infinity, however no
exact solution of the model is given for arbitrary attachment and
detachment rates of both GTP and GDP subunits.

It was thought for a long time that only microtubules were able to
undergo dynamic instability. Recent experiments on single actin
filaments, however, have shown that an actin filament can also
have dynamic-instability-like large length
fluctuations~\cite{fujiwara:02,kuhn-pollard:05}. A similar
behavior was observed in experiments where actin polymerization
was regulated by binding proteins such as ADF/Cofilin
\cite{Michelot-etal:07,roland-etal:08}. Vavylonis et al.
\cite{vavylonis:05} have studied theoretically actin
polymerization kinetics in the absence of binding proteins. Their
model takes into account polymerization, depolymerization and
random ATP hydrolysis. In their work, the ATP hydrolysis was
separated into two steps: the formation of ADP-Pi-actin and the
formation of ADP-actin by releasing the phosphate Pi. Vavylonis et
al. have reported large fluctuations near the critical
concentration, where the growth rate of the filament vanishes.
More recently Stukalin et al. \cite{kolomeisky:06}
have studied another model for actin
polymerization, which takes into account ATP hydrolysis in a
single step (neglecting the
 ADP-Pi-actin state) and occurring only at the interface between
 ATP-actin and ADP-actin (vectorial model) or at a random location
 (random model). This model too shows large fluctuations near
the critical concentration, despite the differences mentioned
above.
Note that both mechanisms (vectorial or random) are still considered
 since experiments are presently not
 able to resolve the cap structure of either microtubule or actin
 filaments.

In this paper, we study the dynamics of a single filament, which
can be either an actin filament or a microtubule, using simple
rules for the chemical reactions occurring at each site of the
filament. The advantage of such a simple coarse-grained
non-equilibrium model is that it  provides insights into the
general phenomenon of self-assembly of linear fibers. Here, we follow the model for
the growth of an actin filament developed in
ref.~\cite{kolomeisky:06}. We describe a new dynamical phase of this
model, which we call the bound phase by analogy with two-state
models of microtubules \cite{Leibler:93}. The characterization of
this bound phase is particularly important, because experimental
observations of time-independent filament distribution of actin or
microtubules \cite{fujiwara:02} are likely to correspond to this
phase. In addition, we analyze the dynamic instability with
 this model. We think that dynamic instability is not a
specific feature of microtubules but could also be present in
actin filaments. We argue that one reason why dynamic instability
is less often observed with actin than with microtubules has to do
with the physical values of some parameters which are less
favorable for actin than for microtubules. This conclusion is also
supported by the work of Hill in his theoretical study of actin
polymerization \cite{hill:85,hill:86}. In these references, a
discrete site-based model for a single actin filament with the
vectorial process of hydrolysis is developed, which has many
similarities with our model.

In short, the model studied in this paper presents three
dynamical phases, which are all non-equilibrium steady states: a
bound phase (phase I), where the average cap length and the filament
length remain constant with time, an intermediate
phase (phase II), where the average cap length remains constant
and the filament grows linearly with time, and a
rapidly growing phase (phase III) where the cap and the filament both grow
linearly with time. The phases II and III were already present in the
study of ref.~\cite{kolomeisky:06}, but phase I was not analyzed there. 
Thus the description of the main features of phase I (such as the average length, the 
distribution of lengths) is one of main results of this paper.

In addition, we discuss how GTP/ATP hydrolysis affects
force-velocity curves and we characterize the large fluctuations
of the filament, by calculating the time needed for the cap to
disappear in phase I and II as well as the time needed for the
complete filament to reach a length of zero ({\it i.e.} to
collapse) for the first time in phase I. Due to the simplicity
of the model, we are able to obtain exact
expressions for all these quantities. We also test these results using Monte
Carlo simulations.

\section*{Model}
\begin{figure}
\includegraphics[scale=0.35]{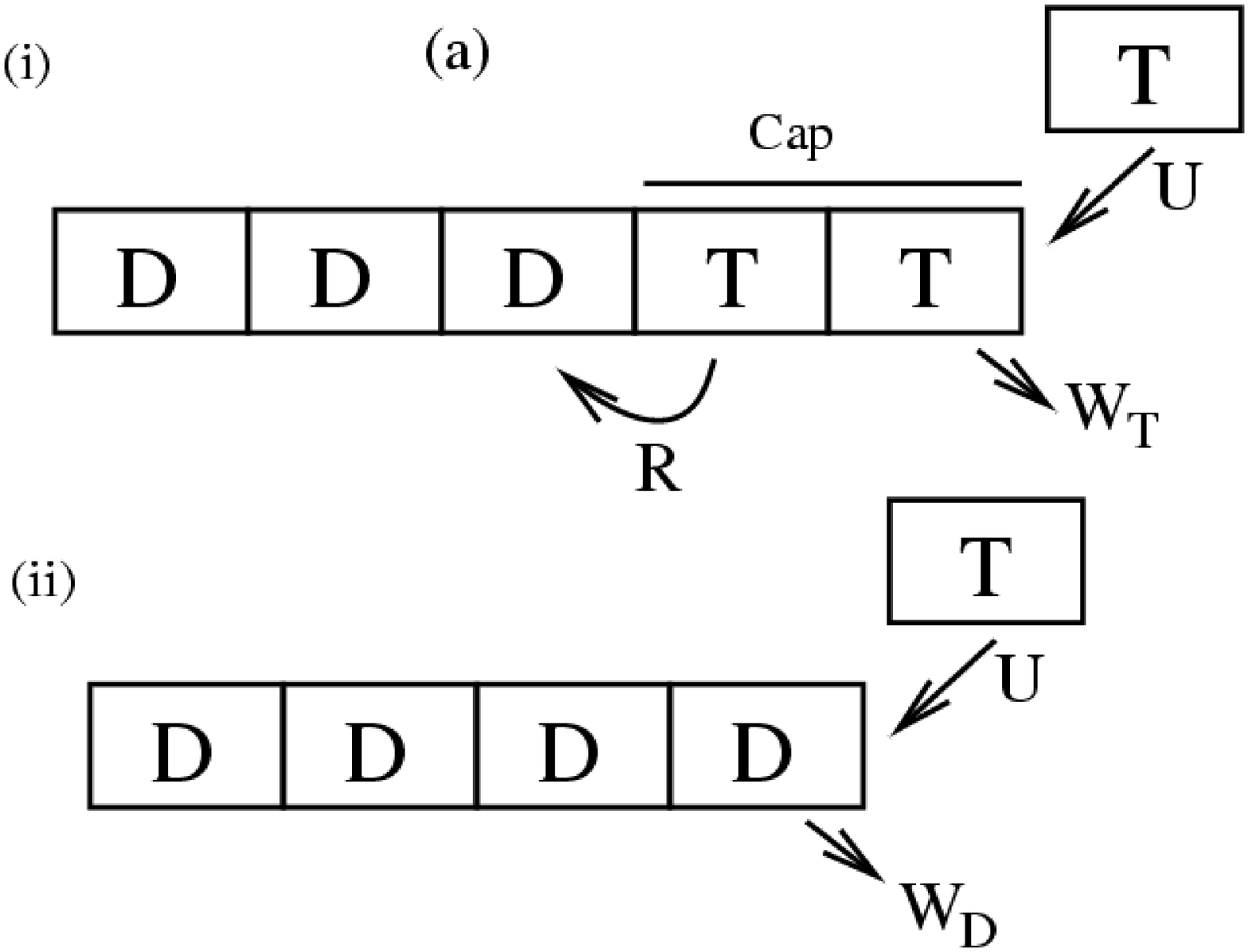}
\includegraphics[scale=0.4]{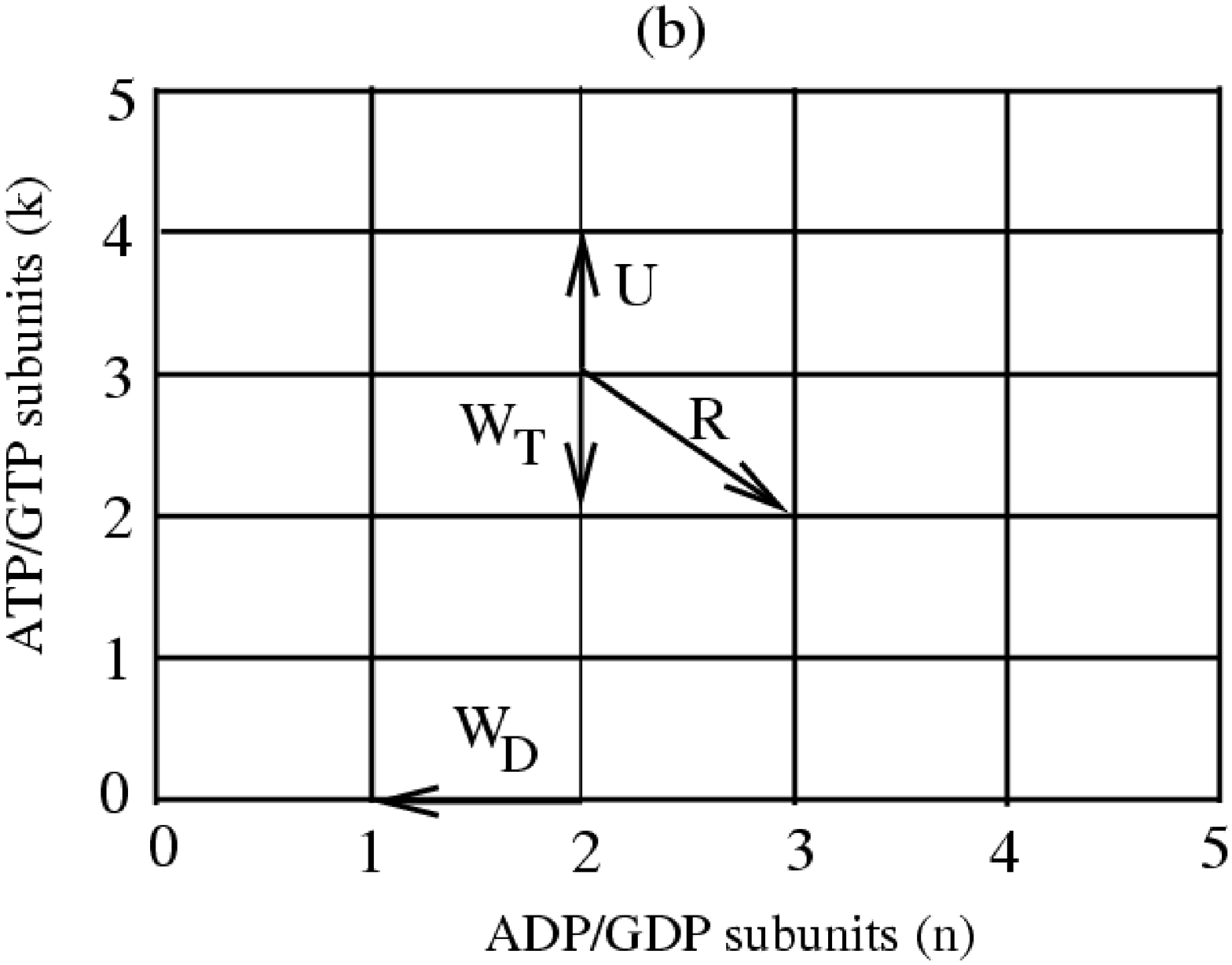}
\caption{\label{fig-model-rates} (a) Schematic diagram showing
addition with rate $U$, removal with rates $W_T$ and $W_D$, and
hydrolysis with rate $R$: (i) for the case where the cap length
is non-zero and (ii) for the case where the cap length is zero. T
stands for ATP (GTP) bound actin (microtubule) subunits while D stands for hydrolyzed
ADP(GDP) subunits. At the T-D interface shown in (i), the hydrolysis
occurs with a rate $R$. (b) Equivalent representation of the
model as a biased random walk in the upper quarter 2D plane with
rates $U$ to go up(north), $W_T$ to go down (south), $R$ to go
south-east and $W_D$ to go west. The $W_D$ move is only possible
along the $k=0$ (southern boundary) line. }
\end{figure}

We study a model for the dynamics of  
of single actin or microtubule filaments
taking into account ATP/GTP hydrolysis. Our model is very much in the
same spirit as that of ref.~\cite{kolomeisky:06} and has also several common features with the 
Hill et al. model \cite{hill:85,hill:86}. We assume that polymerization
occurs, for actin, via the
addition of single ATP subunit (GTP subunit for microtubule), at
the barbed end (plus end for microtubule) \cite{howard-book} of
the filament. We assume that the other end is attached to a wall
and no activity happens there. Let $U$ and $W_T$
  be the rates of addition and removal of ATP/GTP subunits respectively,
  which can occur only at the filament end.
  The subunits on the filament can hydrolyze ATP/GTP and
   become ADP/GDP subunits with a rate $R$. We assume that this process
can occur only at the interface of ATP-ADP or GTP-GDP subunits.
   This corresponds to the vectorial model of
hydrolysis, which is used in the Hill et al. model
\cite{hill:85,hill:86}.
Once
the whole filament is hydrolyzed, the ADP/GDP subunit at the end
of the filament can disassociate with a rate $W_D$. The addition,
removal and hydrolysis events are depicted in
Fig.~\ref{fig-model-rates}. We denote by $d$ the size of a
subunit.

This model provides a simple coarse-grained description of the
non-equilibrium self-assembly of linear fibers. More sophisticated approaches are
possible, which could include in the case of actin, for instance,
additional steps in the reaction such as the conversion of ATP
into ADP-Pi-actin or the possibility of using more than one rate
for the addition of ATP-subunits. It is also possible to
extend our model to include growth from both ends of the filament
rather than from a single end, as discussed in
ref.~\cite{kolomeisky:06}. Another feature of actin or microtubule
filaments, which we leave out in our model, is that these fibers
are composed of several protofilaments (2 for actin and typically
13 for microtubules). In the case of actin, it is reasonable to
ignore the existence of the second protofilament due to strong
inter-strand interactions between the two protofilaments
\cite{kolomeisky:06}. In fact, we argue that the model with a
single filament can be mapped to a related model with two
protofilaments under conditions which are often met in practice.
Indeed the mapping holds provided that the two protofilaments are
strongly coupled, grow in parallel to each other and are initially
displaced by half a monomer. The two models can then be mapped to
each other provided that $d$ is taken to be half the actin monomer
size $d=5.4$nm$/2=2.7$nm. This mapping suggests that many
dynamical features of actin should already be present in a model
which ignores the second protofilament. Similarly, microtubules
may also be modeled using this simple one-filament model, in a
coarse-grained way, provided $d=8$nm$/13=0.6$nm is equal to the
length of a tubulin monomer divided by 13, which is the average
number of protofilaments in a microtubule \cite{howard-book}.
Keeping in mind the fact that the present model is applicable to both actin
and microtubules, we use a terminology appropriate to actin to
simplify the discussion in the rest of the paper.

The actin filament dynamics is studied in terms of two variables
$n$, the number of ADP subunits, and $k$, the number of ATP subunits,
as shown in figure \ref{fig-model-rates}. The dynamics of this system
may be represented as a biased random walk in the upper quarter 2D plane $(n,k)$.
For instance, the addition of one ATP subunit with rate $U$ corresponds to a move
in the upward direction. The removal of ATP subunits with $W_T$ corresponds to
a move in the downward direction. The hydrolysis of an ATP subunit results in
an increase in $n$ and decrease in $k$, both by one unit, which corresponds
to a move in the diagonal direction as shown in the figure. The removal of ADP subunits can
happen only when the cap is zero and therefore corresponds to a leftward move along the $k=0$ line.
Let $P(n,k,t)$ be the probability of having $n$ hydrolyzed
ADP subunits  and  $k$ unhydrolyzed ATP subunits at time $t$, such
that $l=(n+k)d$ is the total length of the filament.  It obeys the
following master equation: For $k>0$ and $n \ge 0$ we have \be
\frac{dP(n,k,t)}{dt}=U P(n,k-1,t)+W_T P(n,k+1,t)+R
P(n-1,k+1,t)-(U+W_T +R)P(n,k,t). \label{pnk} \ee When $n=0$ in Eq.~\ref{pnk},
$P(-1,k+1,t)$ is set equal to zero.

For $k=0$ and $ n \ge 1$ we
have, \be \frac{dP(n,0,t)}{dt}=W_D P(n+1,0,t)+W_T P(n,1,t)+R
P(n-1,1,t)-(U+W_D)P(n,0,t). \label{pn0} \ee If $k=0$ and $n=0$, we
have \be \frac{dP(0,0,t)}{dt}= W_T P(0,1,t)+W_D P(1,0,t) -
UP(0,0,t). \label{p00} \ee The sum of the probabilities is
normalized to $1$ such that \be \sum_{n=0}^{\infty} \sum_{k=0}^{\infty}
P(n,k,t)=1. \ee
We define the following generating functions
\bea
F_k(x,t) &= &\sum_{n \ge 0} P(n,k,t) x^n, \label{fkdefine} \\
H_n(y,t) &= &\sum_{k \ge 0} P(n,k,t) y^k, \\
G(x, y,t) &= &\sum_{n \ge 0} \sum_{k \ge 0}
P(n,k,t) x^n y^k \label{Glg}
 \eea

All quantities of interest can be computed from $G(x,
y,t)$. For instance, the average length of the filament is \be
\langle l \rangle = \left[\langle n \rangle + \langle k \rangle\right]d =  d\left
(\frac{\partial G(x,1,t) }{\partial x}\right)_{x=1}+ d\left(\frac{
\partial G(1,y,t) }{\partial y} \right)_{y=1},
\label{avl}
\ee
where $\langle k \rangle$  is the average number
of the ATP subunits and $\langle n \rangle$ is the average number of
the ADP subunits. The average velocity of the 
filament is
\be
v =\lim_{t \to \infty}\frac{ d \langle l \rangle}{dt}
\label{vdefn},
\ee
and the diffusion coefficient is 
\be
D= \lim_{t \to \infty} \frac{1}{2}
\frac{d}{dt}\left(\langle l^2 \rangle-\langle l \rangle^2\right).
 \label{ddefn}
 \ee
Similarly, the average velocity of the cap is
\be
J= \left[ \lim_{t \to \infty} \frac{ d \langle k
\rangle}{dt} \right] d
\label{Jdefn},
\ee
and the diffusion coefficient of the cap is
\be
D_c= \lim_{t \to \infty}\frac{d^2}{2}
\frac{d}{dt}\left(\langle k^2 \rangle-\langle k \rangle^2\right).
\label{dcdefn}
\ee
The solution of Eqs.~\ref{pnk}, \ref{pn0}, and \ref{p00} and the details of the calculations are given in the Appendix A.

\section*{Phase diagram}
Our model for the dynamics of a single filament with
ATP hydrolysis leads to the following steady-state phases: 
a bound phase (phase I), an intermediate phase (phase II) and 
a rapidly growing phase (phase III).
 In phase I, the bound
phase, the average velocity of the filament $v_I$ and the average
velocity of the cap  both vanish. Thus, the average filament and
cap lengths remain constant in the long time limit. In phase $II$,
the filament is growing linearly in time, with a velocity $v_{II}$,
but the average ATP cap length remains constant as a function of time. In phase $III$, the
filament as well the ATP cap are growing linearly in time with a
filament velocity $v_{III}$ and cap velocity $J$. The boundary
between phases I and II is the curve of equation $v_{II}=0$, and
the boundary between phase II and III is the curve of equation
$J=0$.

We have carried out simulations of the dynamics of the length of
the filament, using the Gillespie algorithm \cite{gillespie:77}.
According to this algorithm, the time to the next on-, off-, or
hydrolysis-event is computed stochastically at each
step of the simulation. We find that our simulation results
 agree with the exact calculations.

\subsection* {The Bound phase (Phase I)}
\begin{figure}
\includegraphics[scale=0.3]{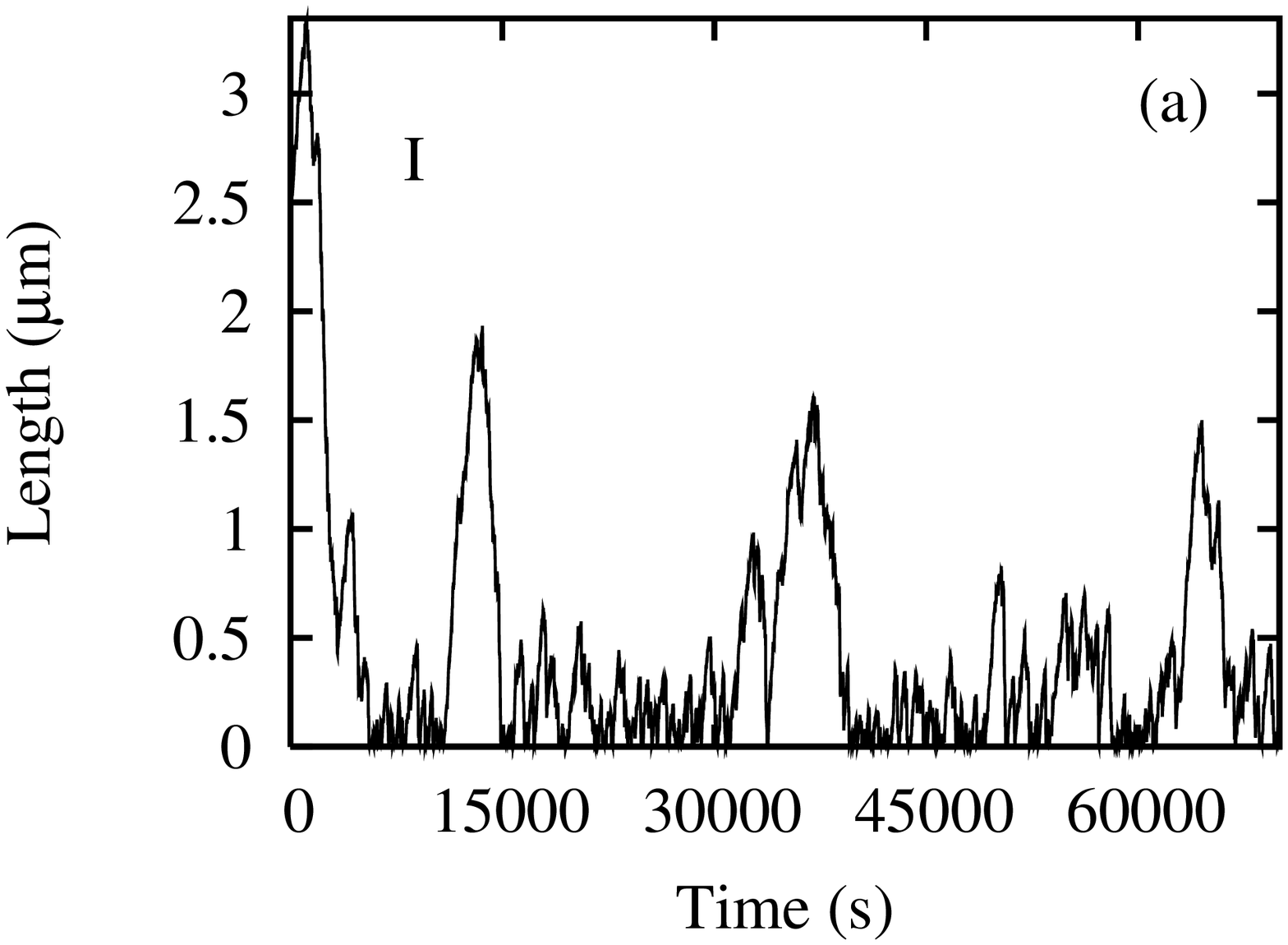}
\includegraphics[scale=0.3]{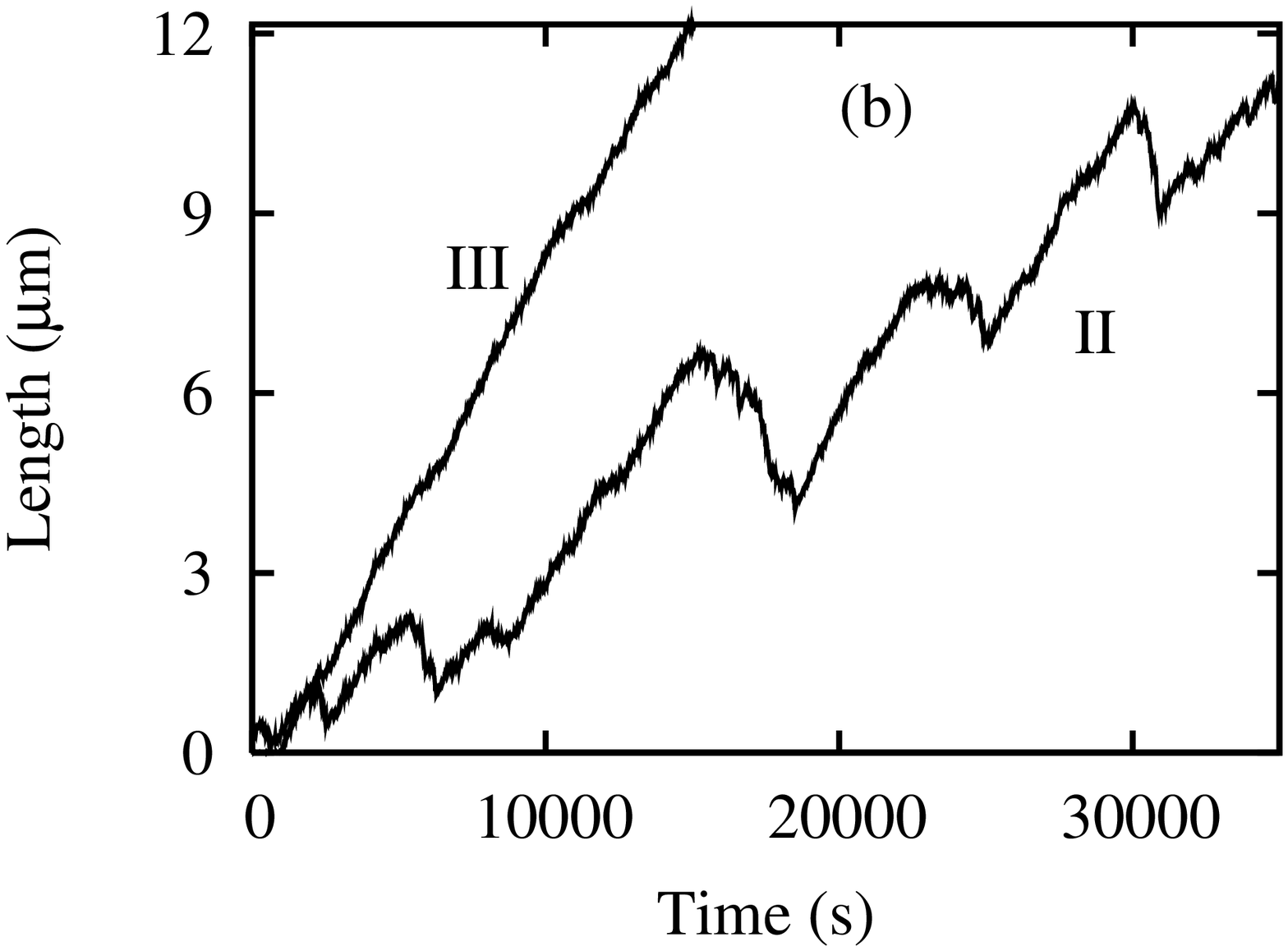}
\caption{\label{fig-Lt1}  Filament length as a function of time for
the three different phases of the model, as computed using Monte
Carlo simulations. The value of the rates which were used are given
in Table~\ref{table-rates}. Figure (a) represents the bound
phase (phase I), where a structure of avalanches in the evolution of the length can be seen.
These avalanches correspond to series of sudden depolymerization events (collapse)
followed by slow polymerization events (rescue).
Figure (b) represents the intermediate
phase (phase II) and the rapidly growing phase (phase III). In the
intermediate phase (II), the dynamics show large length fluctuations as compared
to the rapidly growing phase (III), where the length
fluctuations can hardly be resolved.}
\end{figure}

In the representation of the model as a biased random walk shown
in figure \ref{fig-model-rates}, there is a regime of parameters
for which the biased random walker converges towards the origin.
After some transient time, the random walker enters what we call a
bound steady state, where the motion of the walker is confined to
a bounded region containing the origin. In the representation of
the model as a filament, the filament length fluctuates as
function of time around a time-independent average value $\langle
l \rangle$ and at the same time, the cap length also fluctuates as
function of time around a different time-independent average value
$\langle k \rangle d$. A typical evolution of the total length of
the filament $l(t)$, obtained from our Monte Carlo simulations, is shown in
Fig.~\ref{fig-Lt1}.

%-------------------Table-giving-numerical-value-or-rates------------
\begin{table}
\begin{tabular}{|c|c|c|c|c|c|}
\hline
&$k_0$ ($\mu$m$^{-1}s^{-1}$) & $W_T$($s^{-1}$) &$W_D$($s^{-1}$)& $R$ ($s^{-1}$) & d (nm) \\
\hline
Actin & 11.6 & 1.4 & 7.2 & 0.3 & 2.7 \\
\hline
Microtubule & 3.2  & 24 & 290 & 4 & 0.6 \\
\hline
\end{tabular}
\caption{Numerical estimates of the rates  $W_T$, $W_D$, $R$,
$k_0$ and  subunit length $d$. The parameters for actin are taken
from refs.~\cite{howard-book,kolomeisky:06} and they characterize
the barbed end of an actin filament. The parameters for
microtubule hold similarly for the plus end and at $35^\circ$C.
The rate constant $k_0$ and rate $W_D$ are taken from
\cite{howard-book}, while the other two rates, $W_T$ and $R$, are
deduced from fitting the zero-force velocity data of
ref.~\cite{jason-dogterom:03} and the critical concentration
measurements of ref.~\cite{libchaber_PRE:94}. \label{table-rates}}
\end{table}
%----------------End---Table-giving-numerical-value-or-rates------------
We first discuss the properties of the cap before considering
that of the total length. In the steady state ($t \rightarrow \infty$), $F_k(x=1)$
represents the distribution of cap lengths, as defined in Eq.~\ref{fkdefine}. As shown in the
Appendix A, \be F_k(x=1)=(1-q) q^k, \label{fkx1}\ee
where
\be q=\frac{U}{W_T+R}. \ee
Since $F_0=1-q$, we see that $q$ has the
meaning of the probability of finding a non-zero cap in the steady
state \cite{kolomeisky:06}. We consider for the moment
only the case $q\leq1$, which corresponds to phases I
and II. From $F_k(x=1)$, we find that the average number of cap subunits
is given by
\be
\langle k \rangle = \frac{q}{1-q}, \label{average k} \ee and \footnote{
Note that this expression of $\langle k^2 \rangle$ differs from
that found in ref.~\cite{kolomeisky:06}. This discrepancy, we believe, is probably due to
a misprint in ref.~\cite{kolomeisky:06}.}
 \be
\langle k^2 \rangle = \frac{q+q^2}{(1-q)^2}. \ee As expected,
these quantities diverge when approaching the transition to phase
III when $q\rightarrow1$.  The
standard deviation of the cap length is \be \sigma_{c}^{2} = \langle k^2
\rangle-\langle k \rangle^2=\frac{q}{(1-q)^2}. \ee The relative
fluctuations in the cap size are large since \be
\frac{\sigma_{c}^{2}}{\langle k \rangle^2}=\frac{1}{q} > 1. \ee

\begin{figure}
\includegraphics[scale=0.3]{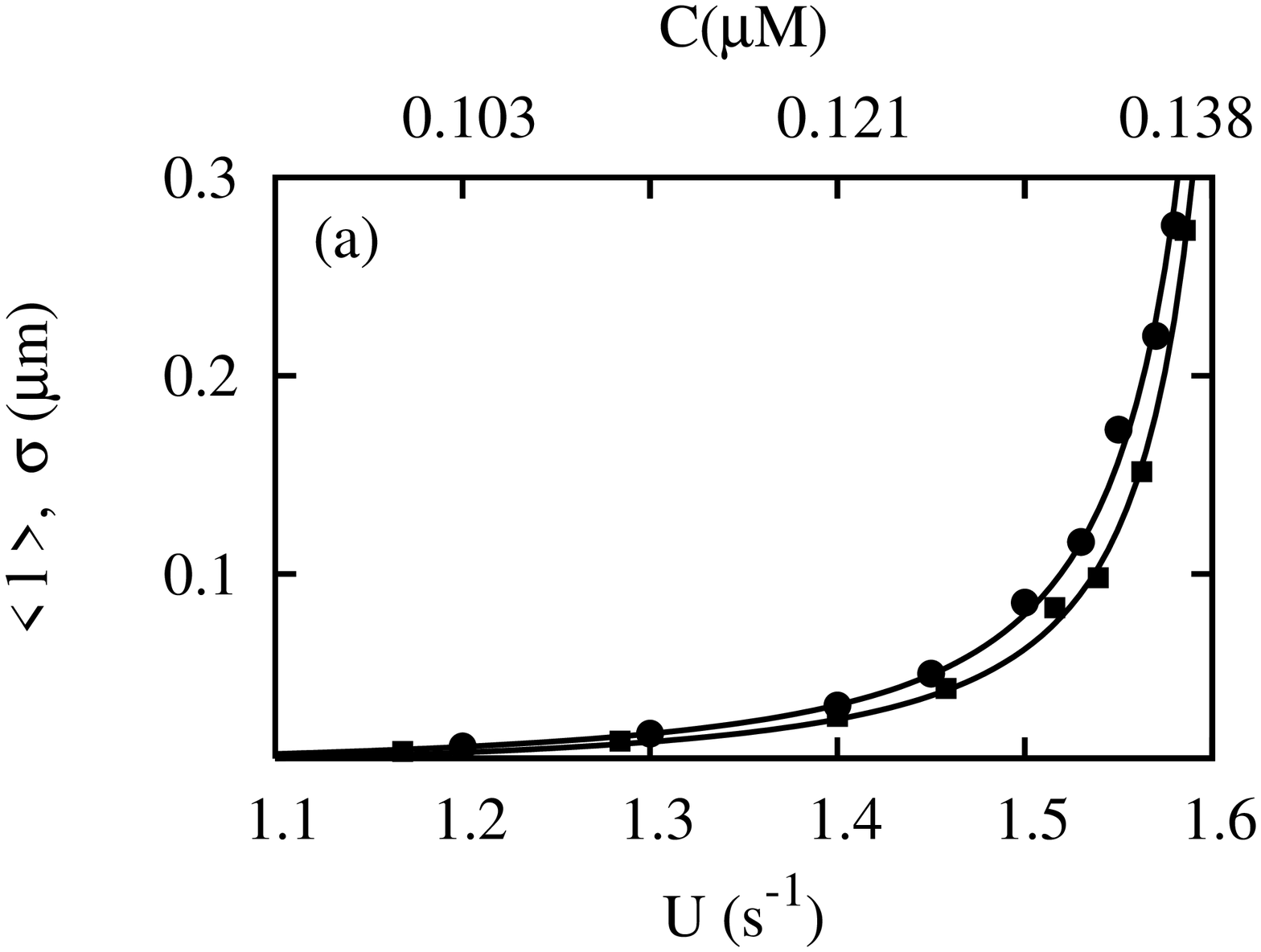}
\includegraphics[scale=0.3]{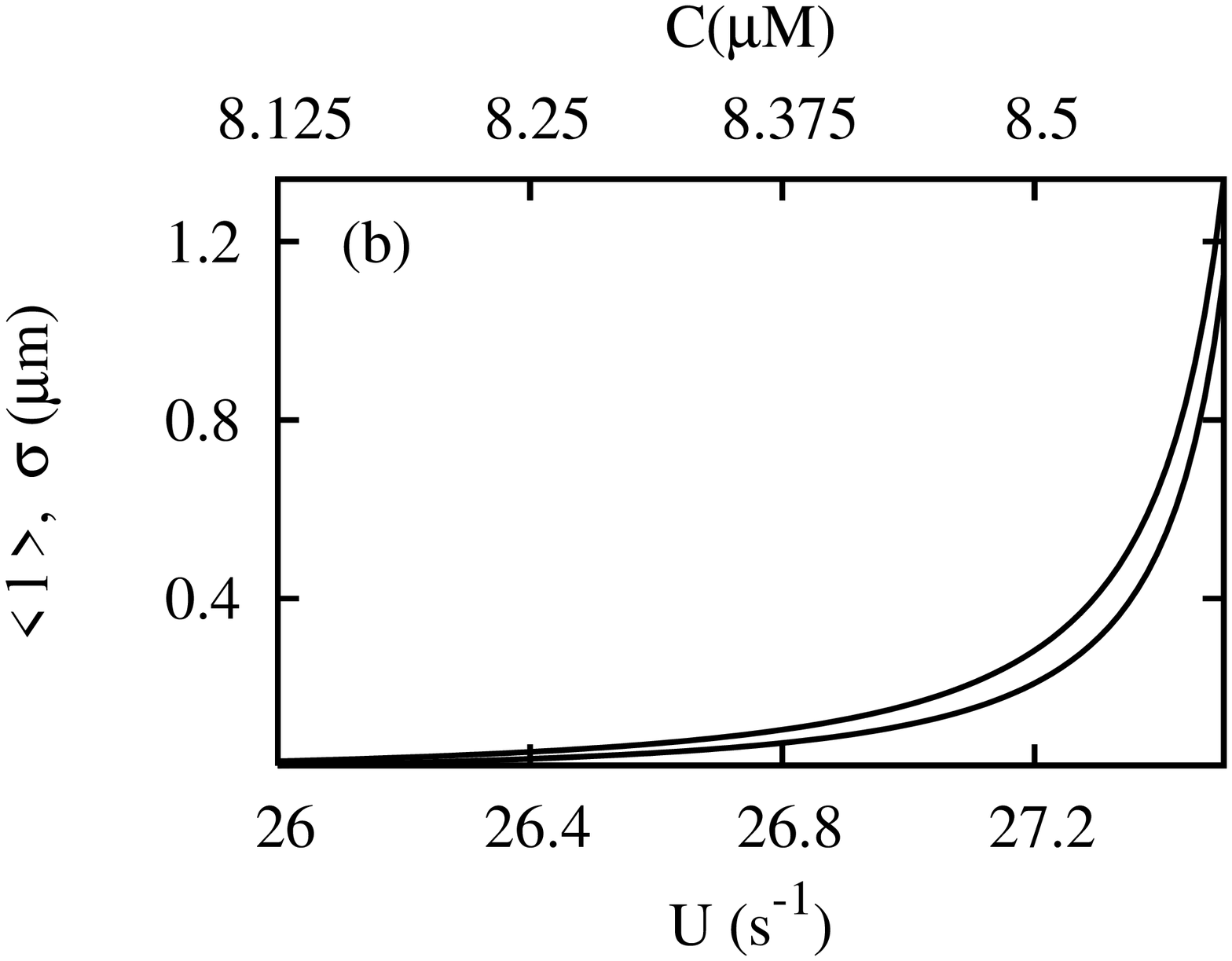}
\caption{\label{fig-LU-bound} Average length $\langle l \rangle$
and its standard deviation $\sigma$  in
the bound phase for (a) actin, and (b) microtubule as a function
of the insertion rate,  $U$(lower x-axis), and of the concentration
of the free ATP/GTP subunits $C$ (upper x-axis). 
The solid lines represent analytical expressions of $\langle l \rangle$
(lower curves) and  $\sigma$ (upper curves). Filled symbols represent values
obtained from simulations for these quantities.
All the curves and symbols are  obtained using values of the rates given in Table.~\ref{table-rates}.}
\end{figure}

We now investigate the overall length of the filament in the
bound phase. This quantity together with the distribution of
length in the bound phase can be obtained from the
time-independent generating function $G(x,y)$. The details of the
calculation of this quantity are given in appendix A. We find
\be G(x,y)= \frac{ \left(W_T +Rx \right) \left(
\frac{W_D}{U}-\frac{W_D+R}{W_T+R} \right) \left(x-1
\right)}{(y-y_{+})[ W_T (y_{-} - 1)x+Rx(y_{-}-x)+W_D y_{-} (1-x) ]},
\label{gxy-final}
 \ee
where $y_{\pm}$ are defined by \be y_{\pm}=\frac{1}{2U} \left(U+W_T+R
\pm \sqrt{(U+W_T+R)^2 -4 U(W_T+Rx)}\right).\ee From the
derivatives of $G(x,y)$, we obtain analytically the average length
$\langle l \rangle$ using Eq.~\ref{avl} as
\bea
\langle l \rangle &=&
\left[\frac{q \left( R^3+W_D R^2+2 R^2 W_T+W_T^2 R + 2
W_D W_T R  + W_D W_T^2 \right)}{ v_{II}(q-1)
(W_T+R)^2}\right] d^2 \nonumber \\
&&
-\left[\frac{q^2 \left( R^2 +2 W_T R+ W_D W_T  \right)}{ v_{II}(q-1)
(W_T+R)}\right] d^2,
\label{lu-eqn}
\eea
where $v_{II}$ is the shrinking velocity (since $v_{II}<0$ in this regime)
of the filament
\be
v_{II}=\left[\frac{U (W_D + R)}{W_T+R} - W_D\right]d.
\label{v2shrink}
\ee
The length $\langle l \rangle$ diverges since $v_{II} \rightarrow 0$
when approaching the transition line between phases I and phase II.
The length $\langle l \rangle$ as given by Eq.~\ref{lu-eqn} is
plotted in Fig.~\ref{fig-LU-bound} for the parameters of table I.
We compare this exact expression with the result of our Monte
Carlo simulations where the average is computed using 1000 length
values taken from different realizations. Excellent agreement is
found with the analytical expression of Eq.~\ref{lu-eqn}.
According to a simple dimensional argument, the average length
$\langle l \rangle$ should scale as $-D_{II}/v_{II}$, where
$D_{II}$ and $v_{II}$ are the diffusion coefficient and velocity
of phase II. We find that this scaling argument actually
holds only close to the transition point between phase I and II. On the
boundary line between phases I and II, the average filament
velocity vanishes and hence the filament length is effectively
undergoing an unbiased ``random walk". In such a case, we expect
that on the boundary line $\langle l^2 \rangle  \sim t$. We have also considered the fluctuations of
$l(t)$ using the standard deviation $\sigma$ defined as \be
\sigma^2= \langle l^2 \rangle - \langle l \rangle^2, \ee for which
an explicit expression can be obtained from $G(x,y)$. In Fig.
\ref{fig-LU-bound}, $\sigma$ is shown as a function of $U$. Note
that $\sigma$ is larger than $\langle l \rangle$, which
corresponds to dynamic-instability-like large length fluctuations.

In the limit $R \rightarrow 0$, ATP hydrolysis can be ignored in
the assembly process. The model is then equivalent to a simple 1D
random walk with rates of growth $U$ and decay $W_T$. In this
case, phases II and III merge into a single growing phase. We find
from Eq.~\ref{lu-eqn} that $\langle l \rangle=Ud/(W_T-U)$, which
diverges as expected near the transition to the growing phase when
$U \simeq W_T$. According to the simple dimensional argument
mentioned above, this length must scale as $-D/v$ in terms of the
diffusion coefficient and velocity of the growing phase
\cite{Leibler-cap:96}. This is the case, since
$D=d^2(U+W_T)/2$ and $v=(U-W_T)d<0$ and thus
$\langle l \rangle=Ud/(W_T-U)$ near the transition point.

\begin{figure}
\includegraphics[scale=0.3]{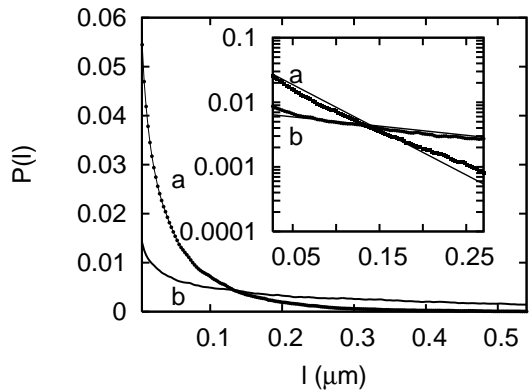}
\caption{\label{fig_pofl} Filament length distribution in the bound phase
for actin. (a) For U=1.5 $s^{-1}$ and (b) for U=1.6 $s^{-1}$ with all other
parameters taken from Table \ref{table-rates}.
The inset shows the same quantities ($l$ vs P(l)) with
$P(l)$ in the log scale. The straight line in the inset is given by Eq.~\ref{eq-pofl}. }
\end{figure}

We have also computed the filament length distribution, $P(l)$  in this phase
using Monte Carlo simulations, as shown in Fig.~\ref{fig_pofl}.
In the inset, we compare the numerically
obtained distribution with the following exponential distribution 
\be
P(l)=P_0 \exp{(-l/\langle l \rangle )},
\label{eq-pofl}
\ee
where $\langle l \rangle$ is given by Eq.~\ref{lu-eqn}.
In this figure, the distribution appears to be close to this exponential
distribution. For any exponential distribution, the standard deviation, $\sigma$ should equal the mean  $\langle l \rangle$. However as seen in figure \ref{fig-LU-bound},
there is a difference between $\sigma$ and $\langle l \rangle$. Hence the distribution
can not be a simple exponential, which could also have been guessed from the fact that the expression of $G(x,y)$ is complicated. The exact analytical expression of the distribution could be
calculated by performing an inverse Z-transform of the known $G(x,y)$.

In the bound phase, experiments with actin in ref.~\cite{fujiwara:02} report an average length
 in the 5-20 $\mu$m range at different monomer concentrations, and experiments with microtubules of ref.~\cite{libchaber_PRE:94} report  a range 1-20 $\mu$m at different temperatures. Neither experiment corresponds precisely to the conditions for which the rates of table \ref{table-rates} are known. Thus a precise comparison is not possible at the moment, although we can certainly obtain with the present approach an average length in the range of microns using the rates of table \ref{table-rates} as shown in figure \ref{fig-LU-bound}.

\subsection*{Intermediate phase (Phase II)}
In the intermediate phase (phase II), the average ATP cap length remains
constant as a function of time, while the filament grows linearly with time. The
presence of this cap leads to interesting dynamics for the
filament. A typical time evolution of the filament length is shown
in figure \ref{fig-Lt1}. One can see the filament switching
between growth (polymerization) and decay (depolymerization) in a
way which is completely analogous to what is observed in the
microtubule dynamics~\cite{Leibler:93}.

In this phase II, the average velocity of the filament is  
\be v_{II} = \left[U -
W_{T} q - W_{D} (1-q)\right]d. \label{velq} \ee
This expression of
$v_{II}$ is the same as Eq.~\ref{v2shrink}
except that now Eq.~\ref{velq} corresponds to the regime where $v_{II}>0$.
The diffusion
coefficient in this phase is \be D_{II}=\frac{d^2}{2}\left[U + W_{T} q + W_{D} (1-q)
+ \frac{2(W_D-W_T)(U+W_Dq)}{W_T+R}\right]. \label{Delq} \ee
The expressions of $v_{II}$ and $D_{II}$ are derived in the appendix A. When $d$ equals to half
the size of an actin subunit, we recover exactly the expressions
of ref.~\cite{kolomeisky:06}.

The transition between the bound phase (I) and the intermediate
phase (II) is delimited by the $v_{II}=0$ curve. When going from
phase $I$ to $II$, the average length in Eq.~\ref{lu-eqn} varies
as $\left(U - W_D\left(\frac{W_T+R}{W_D+R}\right) \right)^{-1}$,
and the variance of the length $\sigma^2$ varies as
$\left(U - W_D\left(\frac{W_T+R}{W_D+R}\right) \right)^{-2}$. The
transition from the intermediate phase II to the rapidly growing
phase III is marked by a similar behavior. The cap length diverges
as $(U-W_T-R)^{-1}$, and the variance of the fluctuations of the
cap length diverges as $(U-W_T-R)^{-2}$.

\subsection*{Rapidly Growing Phase (Phase III)}
In phase III, the length of the ATP cap and that of the filament are
growing linearly with time. Thus the probability of finding a cap
of zero length is zero in the limit $t \rightarrow \infty$, that
is $F_0(x=1)=0$. This also means that the probability of having a
filament of zero length is also zero, {\it i.e.} P(0,0)=0. In this
case, Eq.~\ref{dgdt} reduces to \be \frac{dG(x,y,t)}{dt}=
\left[ U\left(x-1\right) + W_T \left(\frac{1}{y}-1\right) + R
\left(\frac{x}{y}-1\right)\right]G(x,y,t). \ee Using
Eqs.~\ref{vdefn0}, \ref{ddefn0}, \ref{Jdefn0} and \ref{dcdefn0} of Appendix A, one
can easily obtain the following quantities \bea
v_{III} &=& \left[U-W_T\right]d, \label{v3} \\
D_{III} &=& \frac{d^2}{2} \left( U+W_T \right), \\
J&=&\left[U-(W_T+R)\right]d,\label{capv}\\
D_c&=& \frac{d^2}{2} \left(U+W_T+R\right). \eea

Note that these quantities can be obtained from
Eqs.~\ref{velq}-\ref{Delq} by taking the limit $q\rightarrow1$,
which marks the transition between phase III and phase II.
In Fig. \ref{fig-Lt1}, the filament length is plotted as a
function of time. Note that in this phase the velocity and the
diffusion coefficient are the same as those of a filament with no
ATP hydrolysis. The physical reason is that in phase III,
the length of the non-hydrolyzed region (cap) is very large and the
region with hydrolyzed subunits is never exposed.

\section*{Effect of force and actin concentration on active polymerization}
The driving force of self-assembly of the filament is the
difference of chemical potential between bound and unbound ATP
actin subunits. Since the chemical potential of unbound ATP actin
subunits depend on the concentration $C$ of free ATP actin subunits
and on the external applied force $f$, the rates should depend
also on these physical parameters.
In the biological context, this external force corresponds
to the common situation where a filament is pushing against a cell membrane.
For the concentration
dependance, we assume a simple first order kinetics for the
binding of ATP actin monomers given that the solution is dilute in
these monomers. This means that the rate $U$ of binding of ATP actin is
proportional to $C$ while $W_T$, $W_D$, and $R$ should be
independent of $C$~\cite{Peskin_Oster:93,Mogilner_Oster:96,van-doorn:00}. For the force
dependance of the rates, general thermodynamical arguments only
enforce a constraint on the ratio of the rates of binding to that
of unbinding~\cite{hill:81,howard-book}. A simple choice
consistent with this and supported by microtubule experiments
\cite{Dogterom-yurke:97} is to assume that only the binding rate {\it
i.e.} $U$ is force dependent. A more sophisticated modeling of the force dependance of the
rates has been considered 
for instance for microtubules in \cite{kolomeisky_BJ:01}. All the
constraints are then satisfied by assuming that $U=k_0 C
\exp{(-fd/k_BT)}$, with $k_0$, $W_T$, $W_D$, and $R$ all
independent of the force $f$ and of the concentration $C$. We assume
that $f>0$, so that the on-rate is reduced by the application of the force.
The cap velocity in the rapidly growing phase(phase III), given by
Eq.~\ref{capv}, can be written in terms of $f$ and $C$ as \be
J(f,C)= \left[k_0 C e^{-fd/k_BT}-(W_T+R)\right]d. \label{j-cf} \ee
The phase boundary between phase II and phase III is defined by
the curve $J(C,f=f_c)=0$. Equating the cap velocity to zero, we obtain the
characteristic force,
 \be f_c = - \frac{k_B T}{d} \ln{
\left[ \frac{W_T +R}{k_0C} \right] } = - \frac{k_B T}{d} \ln{
\frac{C_0}{C}  }. \ee
where the concentration $C_0$ is defined as
\be
C_0=(W_T+R)/k_0.
\ee
Below $f_c$, the system is in phase III. This is also the point where $q=1$.

The force-velocity relation in the intermediate phase is rewritten, using Eq.~\ref{velq}, as
\be
v_{II}(f,C) = k_0 C e^{-fd/k_BT} \left[ \frac{W_D+R}{W_T+R}
\right]d-W_Dd. \label{vel_II}
\ee
The stall force $f_s$ is by definition the force at which
$v_{II}(f=f_s,C)=0$. From Eq.~\ref{vel_II}, we obtain \be f_s =
-\frac{k_B T}{d}\ln{\left[ \left( \frac{W_T+R}{W_D+R}
\right)\frac{W_D}{k_0 C} \right]}, \ee which can be written
equivalently in terms of the critical concentration of the barbed
end $C_{crit}$ as \be f_s = -\frac{k_B
T}{d}\ln{\left(\frac{C_{crit}}{C} \right)}, \label{fs} \ee where
\be
C_{crit} = C_0 \left( \frac{W_D}{W_D+R}\right) < C_0.
\ee
In the absence of hydrolysis, when $R \rightarrow 0$, we have $C_{crit}=C_0$ and
Eq.~\ref{fs} gives the usual expression of the stall force given in the literature
\cite{howard-book,footer-dogterom:07,Peskin_Oster:93,Mogilner_Oster:96}

\begin{figure}
\includegraphics[scale=0.3]{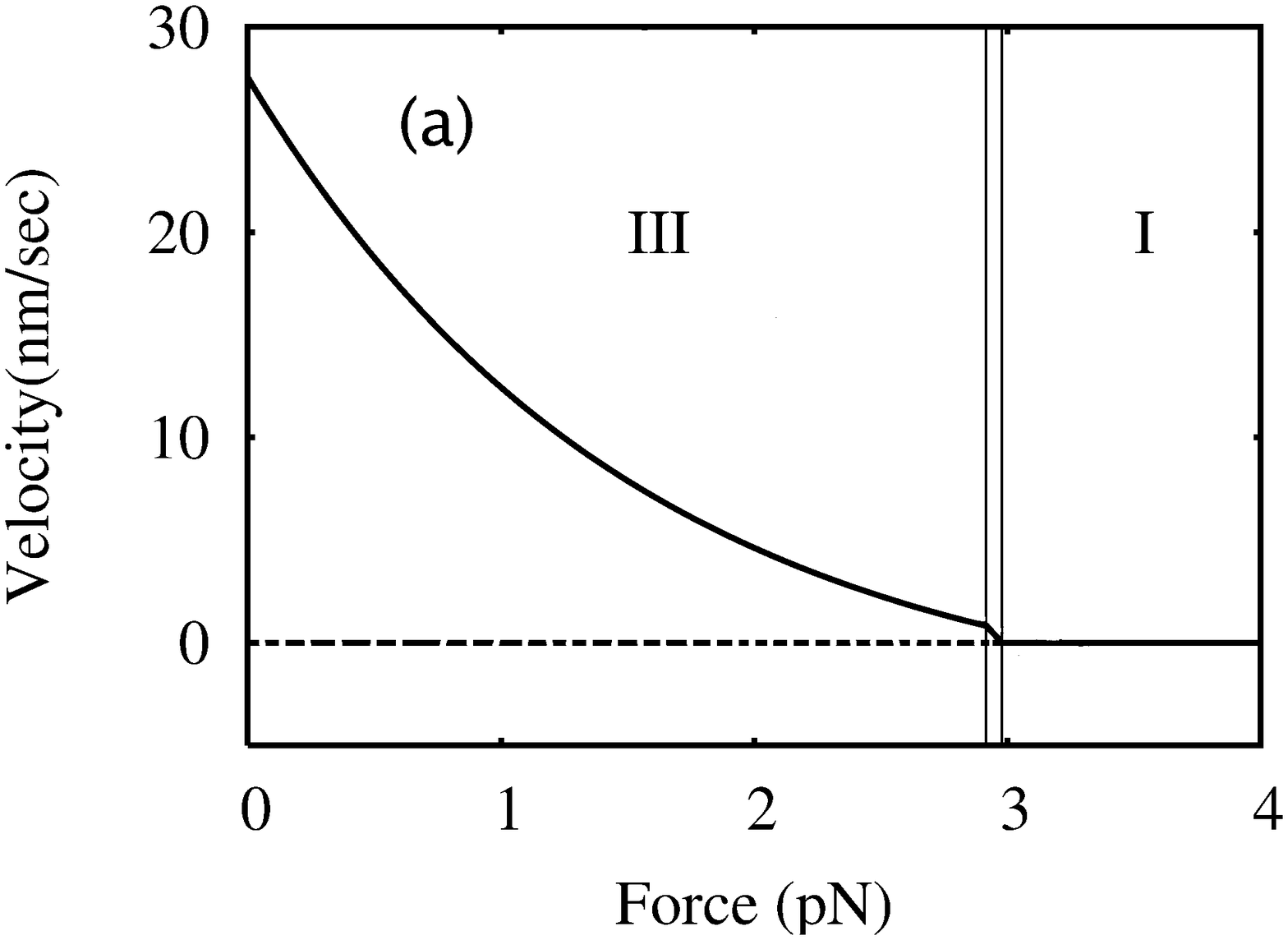}
\includegraphics[scale=0.3]{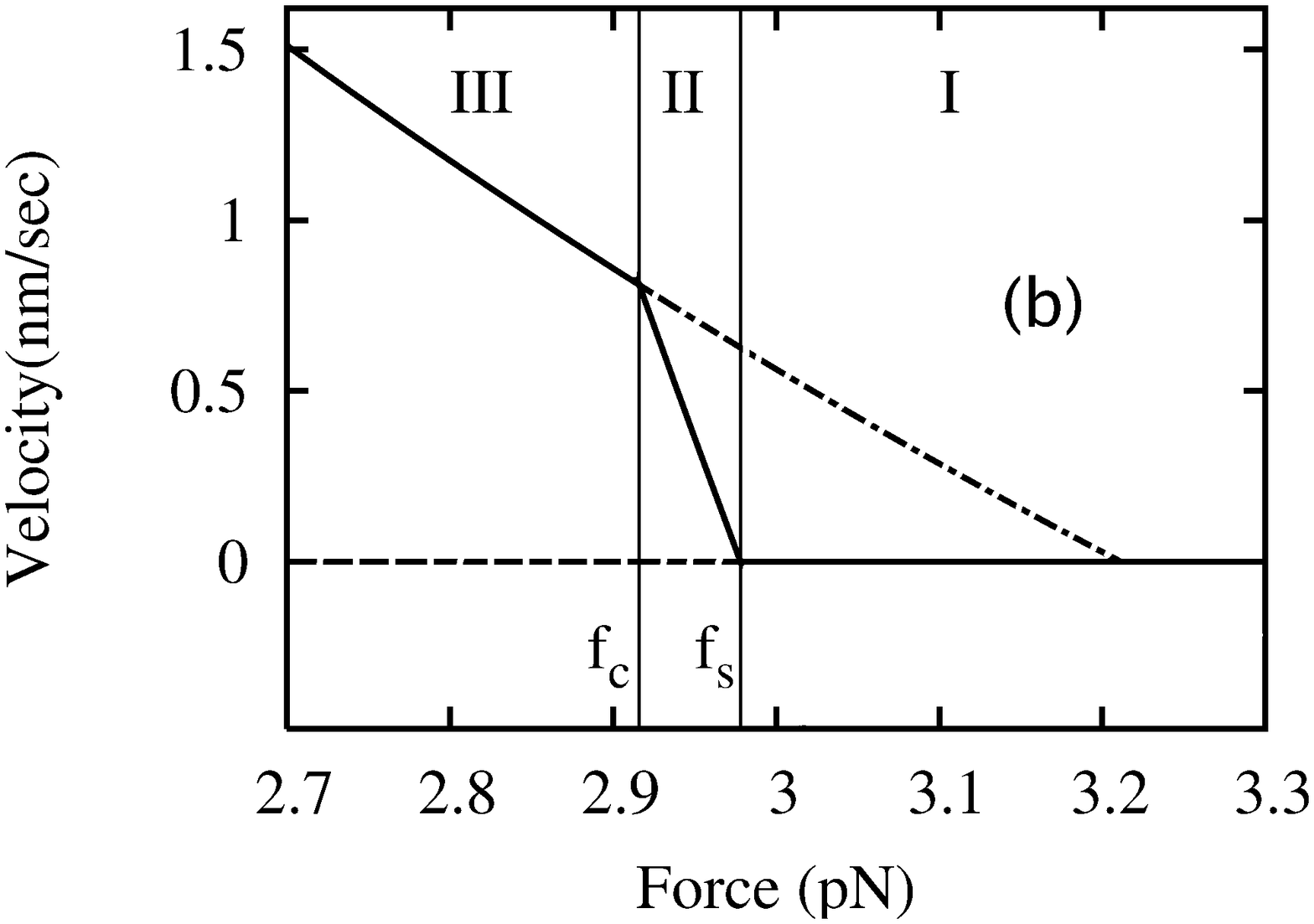}
\caption{\label{fig-fv1} (a) Steady-state force-velocity relation for
a single actin filament shown for $C=1\mu$M (solid curve), 
(b) Zoom of the force-velocity relation near the
stall force.
 The vertical lines represent $f_c$ and $f_s$ as shown in figure (b). For $f<f_c$,
the filament is in the rapidly growing phase (III) and the velocity
is given by Eq.~\ref{v3}. In the bound
phase (I) the velocity is zero. In the intermediate phase the
velocity is given by Eq.~\ref{velq}. The dash-dotted line (figure (b), phase I and II)
is given by Eq.~\ref{v3}, showing that the stall force is higher when ATP hydrolysis is neglected.}
\end{figure}

The velocity of the filament is shown in Fig. \ref{fig-fv1}
with a zoomed in version marked as (b).
This figure shows that for $f<f_c$, the filament is in phase III,
and that there the velocities of the filament with ATP hydrolysis or without are the
same. At $f=f_c$, the force-velocity curve changes its slope, as
shown after the vertical line in figure \ref{fig-fv1} (see the
zoomed-in figure \ref{fig-fv1}b). When the concentration rather than the force is
varied, a similar change of slope is observed at $C=C_0$, which is
accompanied by a discontinuity of the diffusion coefficient
slightly above the critical concentration \cite{kolomeisky:06}.

For $f_c<f<f_s$, the filament is in the intermediate phase, where
the velocities in the presence and in the absence of
ATP hydrolysis differ. The stall force with
ATP hydrolysis is smaller than that in the absence of ATP
hydrolysis. In view of this, a useful conclusion is that it is
important to take into account the ATP hydrolysis for estimating
the velocity of a filament when the force is close to the stall
force.

For $f>f_s$, the velocity of the filament vanishes. It must be
noted that, in this phase, the instantaneous
velocity can be positive or negative, but the average velocity, in the
long time limit, is zero.
Another important point to note is that when the
filament is stalled, ATP is still hydrolyzed. This is
analogous with models of molecular motors containing more than one
cycle \cite{David:PRL07,David:PRE08,lipowsky07}. Including
the chemical cycle of ATP hydrolysis in addition to
the mechanical cycle of addition/removal of subunits is for this reason important
in the context of actin and microtubule models.
One could imagine testing
these predictions on the effect of ATP hydrolysis on
force-velocity relations by carrying out force-velocity
measurements near stalling conditions of abundant ATP or when
ATP is sequestered by appropriate proteins~\cite{MFC_JBC:07}.

\begin{figure}
\includegraphics[scale=0.32]{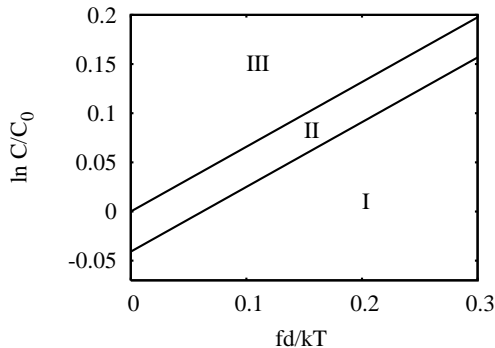}
\caption{\label{fig-muf} Phase diagram as a function of the normalized force
$fd/k_B T$ and of the log of the ratio of the ATP subunit concentration $C$ to the characteristic concentration $C_0$ for actin. The phase diagram shows the bound
phase (phase I), the intermediate phase (phase II), and the rapidly growing phase (phase III). 
The boundary line between
phase II and phase III is the curve
$J=0$ and the boundary line between
phase I and phase II is the curve $v_{II}=0$. }
\end{figure}

All these observations can be summarized in a phase diagram in the
coordinates $f$ and $C$ as shown in Fig.~\ref{fig-muf}.
As shown in Fig.~\ref{fig-muf},  when $C<C_0$, the filament is
either in the intermediate phase (II) or in the bound phase (I).
In this region of the phase diagram, the fluctuations of the
filament length are large as compared to the very small
fluctuations observed in the rapidly growing phase (III). The
large fluctuations observed in phase I correspond to the dynamic instability.

In the case of microtubules, we find as shown in Table~\ref{table-conc},
$C_0 = 8.75 \mu {\rm M}$. This
value is rather large when compared to
typical experimental concentrations $C \simeq 1-10\mu {\rm M}$
for microtubules. Thus microtubules are usually found in phases I and II where
the length fluctuations are large and dynamic instability is commonly observed.

In the case of actin, we find that $C_0=0.147~\mu {\rm M}$.  Typical
experimental actin concentrations are above this estimate,
therefore, at zero force, actin filaments are usually seen in phase III.
This may explain why the dynamic
instability is rarely seen in actin experiments with pure actin.
However, at small concentrations close to the critical concentration,
large length fluctuations have been observed for
actin \cite{fujiwara:02}.

%Table-critical-f-c---
\begin{center}
\begin{table}
\begin{tabular}{|c|c|c|c|c|}
\hline
&$C_0 \, (\mu$M) & $C_{crit} \, (\mu$M) & $f_c$ (pN) & $f_s$ (pN)\\
\hline
Actin & 0.147  & 0.141 &2.916 (at 1 $\mu$M) &2.978  (at 1 $\mu$M)  \\
\hline
Microtubule & 8.75   & 8.63 & 5.65 (at 20 $\mu$M) & 5.74 (at 20 $\mu$M) \\
\hline
\end{tabular}
\caption{Estimates of the characteristic concentrations $C_0$ and $C_{crit}$ at zero force, and 
of the characteristic forces $f_c$ and
        $f_s$ at a concentration of
1 $\mu$M for actin and 20 $\mu$M for microtubule using the rates of Table~\ref{table-rates}.
\label{table-conc}}
\end{table}
\end{center}
%End---Table-giving-numerical-value-or-rates------------

When discussing the effect of force on a single actin or
microtubule filament, one important issue is the buckling of the
filament. Since actin filaments have much smaller persistence
length $l_p$ than microtubules, actin filaments buckle easily
under external force. Our approach is appropriate to describe
experiments like that
 of ref \cite{footer-dogterom:07}, where very short
 actin filaments are used. The length of the filaments must be
 smaller than the critical length for buckling under a force $f$.
 This length can be estimated as $l_{b}= \pi
  \sqrt{\kappa/f}$ with a hinged boundary condition,
  where $\kappa =l_p k_BT$. Taking $l_p \approx 17 \mu$m,
  we estimate $l_b \approx 828$nm at $f=1$pN. Our discussion of the force will
  be applicable only for filaments shorter than $l_b$.

\section*{Collapse time}
In this new section, we shall study experimentally relevant
questions such as the mean time required for the ATP cap to
disappear or the mean time required for the whole filament (ATP
cap and ADP subunits) to collapse to zero length. We are
interested in the conditions for which these times are finite.
Below we address these questions.

\subsection*{Cap collapse in phases I or II}
The dynamics of the cap corresponds to that of a 1D biased random
walker with a growth rate $U$ and a decay rate $W_T+R$. Here, we
calculate the mean time $T_k$ required for a cap of initial length
$k$d to reach zero length for the first time. We assume that there
is a bias towards the origin so that $W_T+R>U$. This time $T_k$ is
nothing but the mean first passage time for the biased random
walker to reach $k=0$, starting from an arbitrary site $k$ in
phases I or II. According to the literature on first passage
times, the equation for $T_k$ is \cite{redner-book,book-saaty,karlin-book} : \be U
T_{k+1} + (W_T + R ) T_{k-1} - (U+W_T+R) T_k +1 =0. \label{rel Tk}
\ee When $W_T+R
> U$,  this recursion relation can be solved (see Appendix B) with the
condition that $T_0=0$, and we obtain \be T_k= \frac{k}{W_T+R - U}.
\label{tk-cap} \ee This corresponds to the time the random walker
takes to travel a distance $k$ at a constant velocity $-J$. Note
that the mean first passage time $T_k$ becomes infinite in the
unbiased case when $J=0$ or if the bias is not towards the origin
{\it i.e.} when $W_T+R \leq U$ (which would correspond to an
initial condition in phase III) \cite{redner-book}.

One can also define an average of the mean first passage
time with respect to the initial conditions. Averaging over $k$
and using equation (\ref{average k}), one obtains \be \langle
T_{k}\rangle = \frac{U}{(W_T + R - U)^2}. \label{tk-av}\ee The
same time can be recovered by considering the average time
associated with the fluctuation of the cap: \be\langle
T_{k}\rangle=\frac{ \langle k^2 \rangle d^2}{2D_c}. \ee This time
may be related to the catastrophe rate in the following way. In
ref.~\cite{Leibler-cap:96}, the catastrophe rate is defined as the
total number of catastrophes observed in an experiment divided by
the total time spent in the growing phase. Since the growing phase
ends when the cap disappears for the first time, we interpret
similarly $1/\langle T_{k}\rangle$, as an average collapse
frequency of the cap.

\subsection*{Filament collapse in phase I}
Now we consider the dynamics of the filament length which is
described similarly by a 2D biased random walk converging towards
the origin. Here we investigate the mean time $T_{n,k}$ required
for a filament with an initial state of $n$ ADP subunits and $k$
ATP subunits to reach zero length for the first time with an
initial condition inside phase I. Again, this is the mean first
passage time now in a 2D domain (in the $n-k$ plane, as shown in
Fig.~\ref{fig-model-rates}) to reach the origin ($n=0,k=0$)
starting from an arbitrary $n$ and $k$. This mean first passage time
$T_{n,k}$ obeys the following set of equations
\cite{redner-book}. When $k>0$, for all $n$, the equation is \be U
T_{n,k+1} + W_T T_{n,k-1}+R T_{n+1,k-1} - (U+W_T+R)T_{n,k}+1 = 0.
\label{kgt0} \ee For $k=0$ and $n>0$ we have a special equation
\be U T_{n,1}+W_D T_{n-1,0}-(U+W_D)T_{n,0}+1=0, \label{keq0} \ee
and we also have the condition $T_{00}=0$.

The simplest way to solve these equations is to guess by analogy
with the 1D case that the solution must be a linear function of
$n$ and $k$. This leads to a simple ansatz of the form $T_{n,k}= An
+ Bk$, which in fact gives the exact result as can be shown rigorously. Substituting this
in Eq.~\ref{kgt0} and Eq.~\ref{keq0}, we can solve for unknowns $A$
and $B$. This leads to \be T_{n,k} =
\frac{nd}{-v_{II}}+\frac{kd}{-v_{II}} \left(
\frac{W_D+R}{W_T+R}\right) \label{tnk-ans} \ee where $v_{II}$ is
the velocity of the intermediate phase given by Eq.~\ref{velq}.
Note that $v_{II}<0$ here since the initial condition is within
phase I.

\begin{figure}
\includegraphics[scale=0.4]{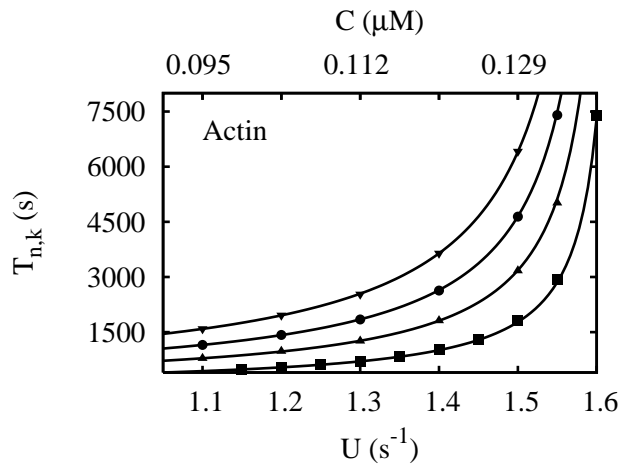}
\caption{\label{fig-Tnk} Mean time taken by a filament of initial
length $(n+k)d$ to collapse to zero length. Curves are given by eq.
\ref{tnk-ans} and points are obtained from a Monte Carlo
simulation for different values of $n$ and $k$. From bottom to top
$(n,k)=(990,10),(750,250),(500,500)$ and $(200,800)$. }
\end{figure}

We first examine some simple particular cases of
Eq.~\ref{tnk-ans}. As we approach the intermediate phase boundary
$v_{II} \rightarrow 0$, $T_{n,k}\rightarrow \infty$ as expected.
When $W_D = \infty$, $T_{n,k}=k/(W_T+R-U)=T_k$, which is the cap
collapse time calculated in the 1D case.
When $W_D = \infty$, the whole filament collapses immediately
after the cap has disappeared for the first time {\it i.e.} after
a time $T_k$. When $R \rightarrow \infty$, ATP subunits
instantaneously become ADP subunits and we obtain another simple result
$T_{n,k}={(n+k)/(W_D-U)}$. We have also compared the prediction of
Eq.~\ref{tnk-ans} with Monte Carlo simulations in
Fig.~\ref{fig-Tnk} and we have found an excellent
agreement.

\begin{figure}
\includegraphics[scale=0.3]{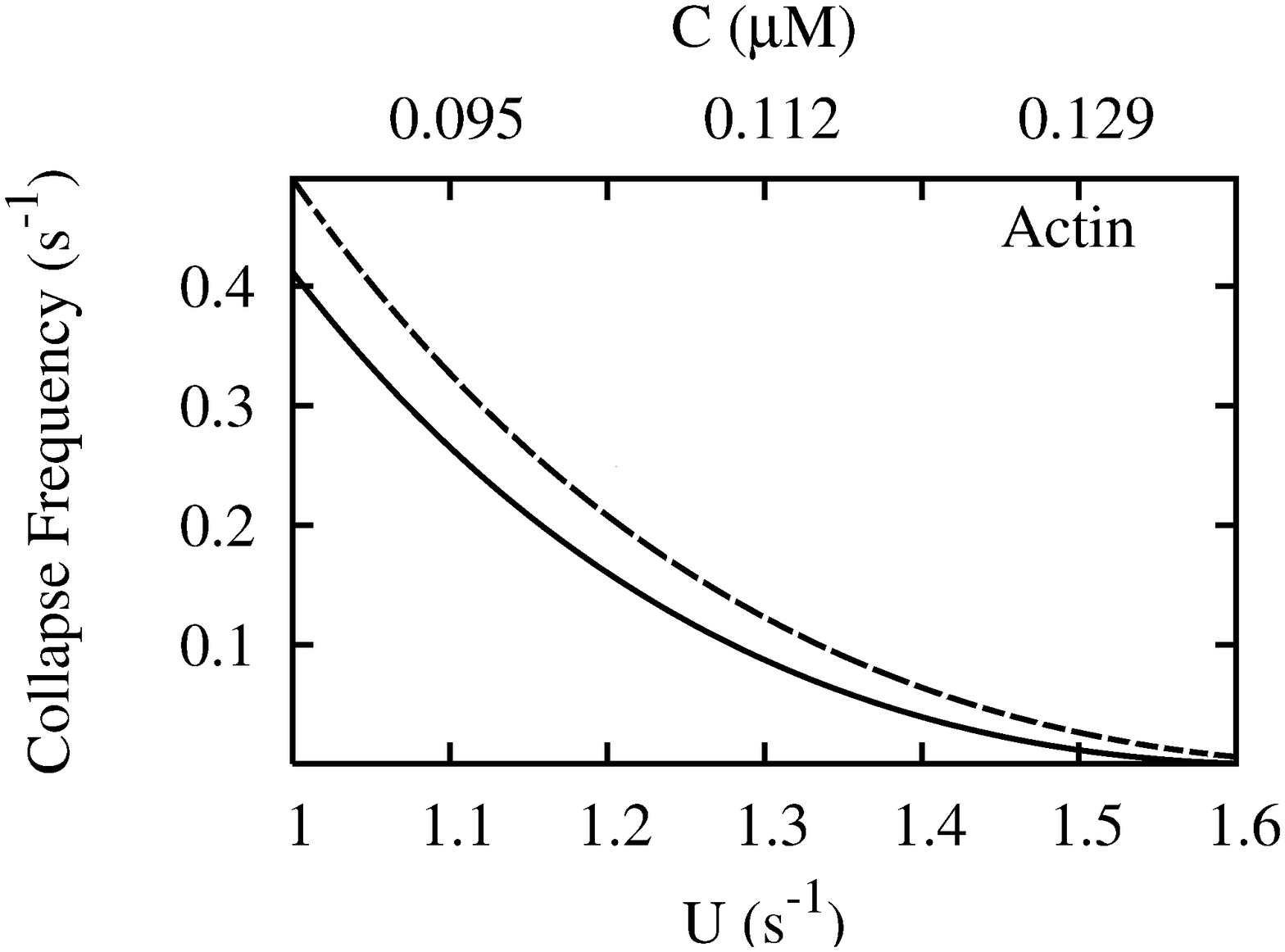}
\includegraphics[scale=0.3]{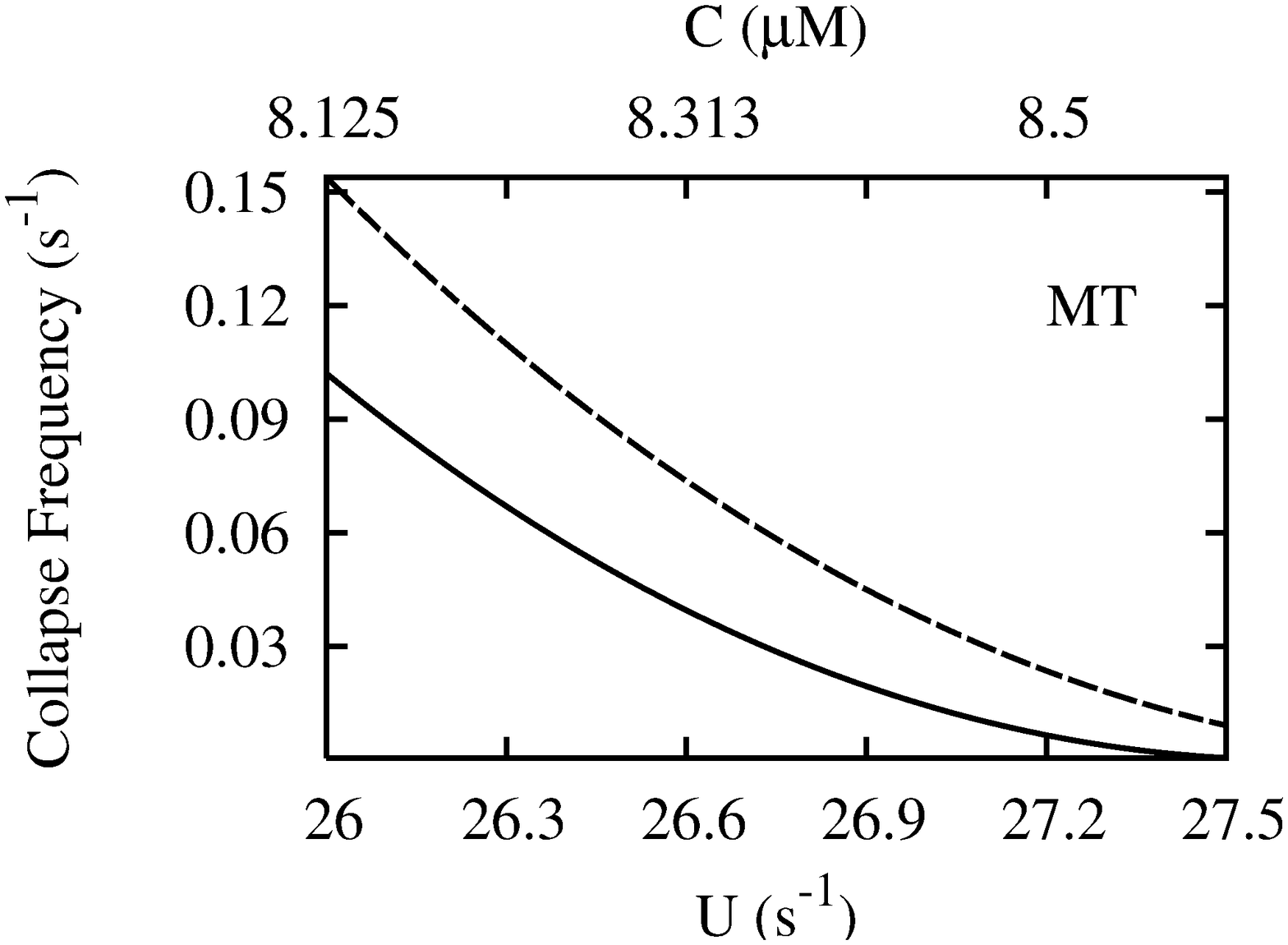}
\caption{\label{fig-collapse-freq}  Collapse frequencies for
actin and microtubule(MT) as function of the rate of addition of monomers $U$ (lower x-axis) and of the
concentration $C$ (upper x-axis): Solid line represents the collapse frequency of the filament
given by, $1/\langle T_{n,k}\rangle$, the inverse of the time given in Eq.~\ref{tnk-av}.
Dashed line represents the  collapse frequency of the cap given by,
$1/\langle T_{k}\rangle$,  the inverse of the time given in Eq.~\ref{tk-av}.}
\end{figure}

We can also define an average of the above mean first
passage time where the average is performed over initial
lengths of cap and un-hydrolyzed region. Averaging
over $k$ and $n$ in Eq.~\ref{tnk-ans} we obtain \be \langle
T_{n,k}\rangle = \frac{U (R^2+W_D R+W_T R+W_D^2)}{ \left(U R-W_D
W_T - W_D R+U W_D\right)^2}. \label{tnk-av} \ee The inverse,
$1/\langle T_{n,k}\rangle$, can be called the collapse frequency
of the filament. The filament collapse frequency and the cap
collapse frequency are shown in Fig.~\ref{fig-collapse-freq} as a
function of $U$ and $C$ for the cases of actin and microtubule
using parameters of Table \ref{table-rates}. Both frequencies are
close to each other because the rate $W_D$ is large compared to
other rates (see Table \ref{table-rates}). This figure also shows
that as the frequency of collapse is increased, the rate $U$
decreases and so the filament length is decreasing, which is
expected \cite{Leibler-cap:96}.
The behavior of the collapse frequency as function of the growing velocity in
the absence of force agrees with \cite{jason-dogterom:03}.
The decrease of the rate of monomer addition
is in practice caused by either the application of a force or a lowering of the concentration.
Thus the application of force may be seen as a general mechanism to regulate the dynamic instability.

In the opposite limit, when $R/W_D \rightarrow 0$ one can also
understand the result physically from the following argument.
When the filament collapses, the first event is the disappearance of the cap
and therefore the first contribution to the collapse time
is  the mean time required  for the cap to
disappear, $T_k$, as obtained from Eq.~(\ref{tk-cap}). Once the cap has
disappeared, assuming that $W_D$ is very large, ADP subunits start
depolymerizing until the next ATP subunit addition takes place. The
mean time needed for an ATP subunit addition to take place is $1/U$.
Once an ATP subunit is added,
one has to wait an average time of $T_{1}$ for the cap to
disappear again. This cycle of ATP subunit addition and
depolymerization repeats many times. The number of times this cycle
occurs, starting with a filament of $n$ ADP subunits, is roughly
$n/(\frac{W_D}{U})$. But one also has
to take into account the increase in ADP subunits as a result of
ATP hydrolysis, which is done by subtracting $RT_1$ from
$n/(\frac{W_D}{U})$.  This leads to the following approximate expression for
$T_{n,k}$
\bea
 T_{n,k} &\simeq& T_k +
\left[\frac{1}{U} + T_1 \right]\frac{n}{(\frac{W_D}{U}-RT_1)}. \\
  & \simeq&
\frac{nd}{-v_{II}}+\frac{kd}{-J} \label{tnk-approx}
\eea where
$J=(U-W_T+R)d$ is the cap velocity in the rapidly growing phase.
This solves Eq.~\ref{kgt0} and Eq.~\ref{keq0} in the limit $R/W_D
\rightarrow 0$  and agrees reasonably well with the Monte Carlo
simulations.

\section*{Conclusions}
In this paper, we have studied a model for the dynamics
of growth and
shrinkage of single actin/microtubule filaments, taking
into account the ATP/GTP hydrolysis which occurs in
the polymerized filament. We find three dynamical phases
with different properties of the ATP/GTP cap and the filament:
a bound phase, an intermediate phase and a rapidly
growing phase. For each phase, we have calculated
the steady-state properties of the non-hydrolyzed
cap and of the filament and we have
investigated the role of an external force (f) applied on
the filament during polymerization and of the
monomer concentration (C), leading to
a f-C phase diagram. We have also calculated
the collapse time which is the time needed for the cap or
the filament to completely depolymerize.

In the bound phase, both the size of the filament and
the size of the cap are finite. It must be noted that
in batch experiments, where the
total (free+bound) concentration of monomers is a constant,
the steady state is always the bound phase.
If the total monomer concentration is large enough,
the filament length is very large and the concentration in the bound phase
will be just below the
critical concentration.  The filaments
exhibit large length fluctuations in the bound phase corresponding
to the dynamic instability observed with microtubules.
An important conclusion is the role of ATP/GTP hydrolysis
in determining
the stall force of the filament: the effect of
hydrolysis always
reduces the value of the stall force. Dynamic
instabilities are in general not observed with
actin filaments when no force is applied to the filament.
However, in recent experiments
 ~\cite{Michelot-etal:07,roland-etal:08},
it has been found that the presence of ADF/Cofilin leads
to a bound phase even at zero force where
actin filaments exhibit large length fluctuations.
In the absence of binding proteins, a natural way to regulate the dynamic instability
and the length of the filament is through the application of force.

The intermediate phase is a phase where the filaments
grow at a constant velocity with a finite ATP/GTP cap.
This is the phase which is in general
observed in a cell.
There are large length fluctuations of the cap in this phase but the length
fluctuations are not as large as the average length.
Thus, there is no true dynamic instability
in this phase but we give predictions for the typical
time of the collapse of the cap. Finally, the growth
phase corresponds to the case where both the filament
and the cap grow at a constant velocity.

One of the limitations of our work is that
we considered a single protofilament. This does
not seem to be an important issue for actin
filaments where the length difference between the two
protofilaments is always small and
of the order of an actin monomer. For microtubules,
the detailed polymerization mechanism seems to be very
complex and this certainly plays an important role on
the way the force is distributed between
protofilaments \cite{Janosi_Flyvbjerg_BPJ:02}. Recent experiments have also
considered the maximum force that can be generated
by a bundle of parallel actin filaments \cite{footer-dogterom:07}.
Our results raise interesting questions about the role
of ATP hydrolysis in this case.

Our work could be extended in several directions.
In the biological context,
actin or microtubule polymerization is regulated by
capping proteins and it would be important to
understand quantitatively the regulation mechanism
and to incorporate them in our model. This is for example
the case for motors of the Kin-13 family that
have been found to interact with microtubules and
induce filament
depolymerization~\cite{julicher_PRL:04}. So far, we 
have mainly considered the polymerization kinetics, but
in general, there is a complex interplay between the mechanical properties of the filaments
and the polymerization kinetics, which we plan to explore in future work.

\section*{Acknowledgment}
We thank G. I. Menon, J. Baudry, C. Brangbour, C. Godr\`{e}che and J. Prost
for useful discussions. We also thank J-M. Luck for illuminating conversations. 
We also acknowledge support from
 the Indo-French Center CEFIPRA (grant 3504-2).

\appendix
\section*{Appendix}

\subsection*{A. Calculations using the generating function approach}

Summing over $n$ and $k$ in Eq.~\ref{pnk} and using equations
\ref{pn0}, \ref{p00} and \ref{Glg}, one obtains
\bea
\frac{dG(x, y, t)}{dt}&=& \left[ U\left(x-1\right) + W_T
\left(\frac{1}{y}-1\right)
+ R \left(\frac{x}{y}-1\right)\right]G(x, y,t) \nonumber \\
&-& \left[ W_T \left(\frac{1}{y}-1\right) + R
 \left(\frac{x}{y}-1\right)+W_D\left(1-\frac{1}{x}\right)
 \right]F_0(x,t) \nonumber\\
&+&W_D\left(1-\frac{1}{x}\right) P(0,0,t).
\label{dgdt}
\eea
Similarly one can also write down equations for $F_k$ and $H_n$.
This equation contains $F_0$, which is coupled to all the $F_k$.
For $k>0$,
\be
\frac{dF_k(x,t)}{dt}= U F_{k-1}(x,t) + \left(W_T +R
x \right)F_{k+1}(x,t) - \left(U+W_T+R \right)F_k (x,t),
\label{fkl}
\ee
and for $k=0$,
\be \frac{dF_0(x,t)}{dt}= \left(W_D
\left(\frac{1}{x}-1\right)-U\right)F_{0}(x,t) +\left(W_T +R
x\right)F_{1}(x,t).
\label{f0l}
\ee
Solving this set of equations we shall derive a formula for
 $G(x,y,t)$. From $G(x,y,t)$, we calculate the
following quantities:

The average length \be \langle l \rangle =
 \left[ \langle n \rangle + \langle k \rangle \right]d
 =  d\left (\frac{\partial G(x,1,t) }{\partial
x}\right)_{x=1}+ d\left(\frac{
\partial G(1,y,t) }{\partial y} \right)_{y=1},
\ee
the velocity of the filament 
\be
v =\lim_{t \to \infty}\frac{ d \langle l \rangle}{dt}=d \lim_{t \to
\infty} \frac{\partial }{\partial
x}\left(\frac{dG(x,x,t)}{dt}\right)_{x=1}
\label{vdefn0}, \ee
and the diffusion coefficient of the filament length
\bea
D&= &\lim_{t \to \infty} \frac{1}{2} \frac{d}{dt}\left(\langle l^2 \rangle-\langle l \rangle^2\right) \nonumber\\
&=&d^2 \lim_{t \to \infty} \left[\frac{1}{2}\frac{\partial^2
}{\partial x^2}\left(\frac{dG(x,x,t)}{dt}\right)
+\frac{1}{2}\frac{\partial }{\partial
x}\left(\frac{dG(x,x,t)}{dt}\right) \right. \nonumber\\&&
\left.-\left(\frac{\partial G(x,x,t) }{\partial
x}\right)\frac{\partial }{\partial
x}\left(\frac{dG(x,x,t)}{dt}\right) \right]_{x=1}. \label{ddefn0}
\eea
The average velocity of the cap is
\be
J=d \lim_{t \to
\infty} \frac{ d \langle k \rangle}{dt}=\lim_{t \to \infty}
\frac{\partial }{\partial
y}\left(\frac{dG(1,y,t)}{dt}\right)_{y=1} \label{Jdefn0},
\ee
and the diffusion coefficient of the cap is \bea D_c&=&
d^2 \lim_{t \to \infty}\frac{1}{2}
\frac{d}{dt}\left(\langle k^2 \rangle-\langle k \rangle^2\right) \nonumber\\
 &=&d^2 \lim_{t \to \infty} \left[\frac{1}{2}\frac{\partial^2 }{\partial
y^2}\left(\frac{dG(1,y,t)}{dt}\right) +\frac{1}{2}\frac{\partial
}{\partial
y}\left(\frac{dG(1,y,t)}{dt}\right) \right. \nonumber \\
& &\left.- \left(\frac{\partial G(1,y,t) }{\partial
y}\right)\frac{\partial }{\partial
y}\left(\frac{dG(1,y,t)}{dt}\right) \right]_{y=1}. \label{dcdefn0}
 \eea

\subsubsection*{Calculation of $F(x=1, t\rightarrow \infty)$ in phases I and II}
In the steady-state ($t\rightarrow \infty$), the cap distribution in phases
I and II becomes time-independent
and hence $(dF_k/dt)_{x=1}=0$.
In this case Eqs. \ref{fkl} and \ref{f0l} can be written, for $k>0$, as
\be
0= U F_{k-1} + \left(W_T +R
 \right)F_{k+1} - \left(U+W_T+R \right)F_k ,
\label{fkl0}
\ee
and for $k=0$,
\be
0=\left(W_T +R
\right)F_{1}-UF_0.
\label{f0l0}
\ee
where we denote for short, $F_k=F_k(x=1,t\rightarrow \infty)$.
The solution of  Eq. \ref{fkl0} is of the
form $F_k=q^k F_0$. If we substitute this back into Eq. \ref{fkl0},
we get a quadratic equation in $q$
\be
(W_T+R)q^2 - (U+W_T+R)q+U=0.
\label{qeqn}
\ee
The two solutions are $q=U/(W_T+R)$ and $q=1$, but we can 
rule out $q=1$ using the normalization condition $\sum_{k=0}^{\infty} F_k=1$. 
In phases I and II,
$W_T+R > U$ and therefore $q<1$. Using the normalization condition, we obtain
\be
F_k=(1-q)q^{k},
\label{fkresult}
\ee
which is Eq.~\ref{fkx1}.

\subsubsection*{Calculation of $G(x,y)$ in the bound phase (phase I) .}
We now explain how to calculate $G(x,y)$ in the bound phase, 
using a technique of canceling apparent poles ~\cite{book-saaty,karlin-book}.
Since
we are interested in the steady-state properties of the bound
phase, the time derivative of $G$ on the left hand side of
Eq.~\ref{dgdt} is zero, which leads to \be G(x,y) =\frac{
F_0(x)\left[ R x (y-x) +W_T x (y-1) -W_D y (x-1)\right] -W_D
P(0,0) y (1-x)}{x \left[-Uy^2 +(U+W_T+R)y -Rx -W_T\right]},
\label{gbound1} \ee where $F_k(x)$ and $P(0,0)$ are unknowns. 
By definition
$$G(x,y)=\sum_{n \ge 0} \sum_{k \ge0 } P(n,k) x^n  y^k,$$
since the $P(n,k)$ are bounded numbers, $G(x,y)$  is an analytic
function for  $0 \le |x| \le 1$ and $0 \le |y| \le 1$. To
guarantee the analyticity of the function $G(x,y)$, the zero of the 
denominator of Eq.~\ref{gbound1},
$$y=y_-= \frac{1}{2U} \left(U+W_T+R - \sqrt{(U+W_T+R)^2 -4 U(W_T+Rx)}\right),$$
 must also be a zero of  the
numerator. This implies that \be
F_0(x)=\frac{W_D P(0,0) y_{-} (1-x)}{R x (y_{-}-x) +W_T x (y_{-}-1)
-W_D y_{-} (x-1)}. \label{F_0} \ee The normalization condition,
namely G(x=1,y=1)=1, then fixes the value of $P(0,0)$ as \be P(0,0)= 1-
\frac{U}{W_D}\left(\frac{W_D+R}{W_T+R}\right). \label{P00} \ee
After substituting Eqs.~\ref{F_0} and \ref{P00} into
Eq.~\ref{gbound1}, we obtain the expression of $G(x,y)$ given in
Eq.~\ref{gxy-final}.

\subsubsection*{Velocity and diffusion coefficient in the Intermediate phase (phase II)}
We recall the definition of $F_k(x,t)$ given in Eq.~\ref{Glg}:
\be F_k(x,t) = \sum_{n \ge 0} P(n,k,t) x^n, \ee
and we recall that $F_k$ with no argument is a short notation for $F_k(x=1,t\rightarrow \infty)$.
From this, we introduce
 \be
 a_k(t) =\left( \frac{\partial F_k(x,t)}{\partial x} \right)_{x=1},
 \label{adefine}
 \ee
 so that $<n(t)>=\sum_{k \ge 0} P(n,k,t) n=\sum_{k \ge 0} a_k(t)$.

By taking a derivative with respect to $x$ in Eqs. \ref{fkl} and \ref{f0l}, one obtains the
equations of evolution of $a_k(t)$: for $k>0$
 \be
\frac{da_k(t)}{dt}= U a_{k-1}(t) + (W_T +R)a_{k+1}(t) - (U+W_T+R)a_k(t)+RF_{k+1}(x=1,t)
\label{akl}
\ee
and for $k=0$
\be
\frac{d a_0(t)}{dt}=  -W_DF_0(x=1,t) -Ua_{0}(t) +(W_T +R)a_{1}(t)+RF_1(x=1,t).
\label{a0l}
\ee
As shown in \cite{kolomeisky:06}, there is a solution of these recursion relations
in the long time limit in the form of
$a_k(t)=M_kt+B_k$, where $M_k$ and $B_k$ are time independent coefficients.
After substituting this equation into Eqs.~\ref{akl}-\ref{a0l}, and separating terms which are time-dependent from terms which are not time-dependent,
one obtains separate recursion relations for $M_k$ and $B_k$.
The recursion relation of $M_k$ is identical to that of $F_k$ obtained in Eq.~\ref{fkl0}-\ref{f0l0}. Using Eq.~\ref{fkresult}, the
solution can be written as $M_k=v_{II}
(1-q)q^k/d=v_{II} F_k/d$. The recursion relation of $B_k$ is
for $k>0$, \be M_k=U B_{k-1} + (W_T
+R)B_{k+1} - (U+W_T+R)B_k+RF_{k+1}, \ee
and for $k=0$,
\be M_0=-W_D F_0 - U B_{0} + (W_T
+R)B_{1} + R F_{1}. \ee
These recursion relations
can also be solved with the result \be B_k=B_0q^k+
\left[\frac{W_D(1-q)}{W_T+R} \right]kq^{k}. \ee

To characterize the intermediate phase, it is convenient
to rewrite the evolution equation for the generating function $G(x,y,t)$ of
Eq.~\ref{dgdt} using the fact that $P(0,0,t\rightarrow \infty)=0$ in this phase,
in the form of an evolution equation for $\tilde{G}(x,t)=G(x,x,t)$ as
\be
\frac{d\tilde{G}(x,t)}{dt}=\alpha(x) \tilde{G}(x,t) + \beta(x,t), \label{DGdt} \ee
where $\alpha(x)=U (x-1) + W_T (1/x-1)$ and
$\beta(x,t)=(1-1/x) (W_T-W_D ) F_0(x,t)$.
With this notation, the velocity defined in Eq.~\ref{vdefn0} is
\be
v =d \lim_{t \to
\infty} \frac{\partial }{\partial
x}\left(\frac{d\tilde{G}(x,t)}{dt}\right)_{x=1}=d \left[ \alpha'(1)+\beta'(1,t\rightarrow \infty) \right], \ee
where the prime denotes derivatives with respect to $x$.
Substituting the expressions of $\alpha(x)$ and $\beta(x,t)$ into this equation,
it is straightforward to obtain
the velocity $v=v_{II}$ characteristic of the intermediate phase which is Eq.~\ref{velq}.
Similarly, using Eq.~\ref{ddefn0} and Eq.~\ref{DGdt}, the diffusion coefficient can be written as
\be
D = \frac{d^2}{2} \lim_{t \to \infty} \left[ \alpha''(1) + \beta''(1,t)
+ \alpha'(1) + \beta'(1,t) - 2 \tilde{G}'(1,t)\beta'(1,t) \right],
\label{simple D} \ee
where
\bea
\tilde{G}^{\prime}(1,t) &=& \left(\frac{\partial \tilde{G}(x,t)}{\partial x }\right)_{x=1}=
\langle n(t) \rangle + \langle k \rangle, \nonumber \\
                        &=& vt + \frac{1}{1-q} \left[ B_0 + \frac{U+W_D q}{W_T+R} \right].
\label{simple G'} \eea
After substituting Eq.~\ref{simple G'} into Eq.~\ref{simple D} and simplifying, 
the terms linear in time and the term containing the unknown parameter $B_0$ 
 cancel out in the expression of the diffusion coefficient, and we finally obtain $D=D_{II}$
which is given in Eq.~\ref{ddefn0}.

 \subsection*{B. Cap collapse time $T_k$}
In order to solve the recursion relation for $T_k$ of Eq.~\ref{rel
Tk}, we perform a $z$ transformation defined by \be
\tilde{T}(z)=\sum_{k\ge0}T_k z^{-k}.  \ee After using the initial
condition $T_0=0$, we obtain \be \tilde{T}(z) =\frac{ (1+U T_1
(1-z))z}{U(1-z)^2(q^{-1}-z)} \label{tdz} \ee
 where $q^{-1}= (W_T+R)/U$. By definition,
 $\tilde{T}(z)$ is analytic for all values of $|z|>1$.
 Since we are interested in the case $W_T+R >U$,
 the numerator in Eq.~\ref{tdz} must vanish to ensure that
 $\tilde{T}(z)$ is analytic at $z=q^{-1}$.
 This condition determines the unknown $T_1=1/(W_T+R-U)$.
 Now $T_k$ can be obtained by an inverse Z transform as
 \be
 T_k=\frac{1}{2 \pi i} \oint\tilde{T}(z)z^{k-1}dz=
 \frac{1}{2 \pi i} \oint \frac{z^k dz}{U(1-z)^2(q^{-1}-1)} = \frac{k}{W_T+R-U}.
 \ee

% closing statement, nothing below matters
\end{document}